\documentclass[aps,twocolumn,prd,keywords,showkeys,showpacs,preprintnumbers,nofootinbib,10pt]{revtex4-2}
\usepackage[utf8]{inputenc}
\usepackage[T1]{fontenc}
\usepackage[margin=2cm]{geometry} 
\usepackage{amsmath}%
\usepackage{amssymb}%
\usepackage{amsfonts}%
\usepackage{amsthm}%
\usepackage{mathtools}%
\usepackage{mathrsfs}%
\usepackage[caption=false, singlelinecheck=false,
justification=raggedright]{subfig}%
\usepackage{xspace}%
\usepackage{txfonts}%
\usepackage{multirow} 
\usepackage{array}%
\usepackage{booktabs}
\usepackage[normalem]{ulem}%
\usepackage{graphicx}%
\usepackage{rotating} 
\usepackage{soul}%
\usepackage{slashed}%
\usepackage{nicefrac}
\usepackage{paralist}

\usepackage[colorlinks, urlcolor=black, citecolor=black, linkcolor=black]{hyperref}%

\newcommand{\modulus}[1]{\left| #1 \right|}

\newcommand{\overbar}[1]{\kern 0.2em\overline{\kern -0.2em #1}{}\xspace}

\allowdisplaybreaks

\begin{document}
\title{Methodology to determine the spin-parity of muon-philic \texorpdfstring{$\boldsymbol{X}$}{X} boson in \texorpdfstring{$\boldsymbol{J/\psi \rightarrow \mu^- \mu^+ X}$}{J/psi --> mu- mu+ X} decay}

\author{Manimala Mitra}
\email[Email at: ]{manimala@iopb.res.in}
\affiliation{Institute of Physics, Sachivalaya Marg, Bhubaneswar 751005, India}
\affiliation{Homi Bhabha National Institute, BARC Training School Complex, Anushaktinagar, Mumbai 400094, India}

\author{Dibyakrupa Sahoo}
\email[Email at: ]{dibyakrupa.s@iopb.res.in}
\affiliation{Institute of Physics, Sachivalaya Marg, Bhubaneswar 751005, India}
\affiliation{Homi Bhabha National Institute, BARC Training School Complex, Anushaktinagar, Mumbai 400094, India}

\date{\today}

\begin{abstract}
The present anomaly in muon anaomalous magnetic moment can be explained by the
presence of a muon-philic $X$ boson which could be a scalar particle or a vector
particle with mass less than twice the mass of muon. The muon-philic $X$ boson
could interestingly not be a parity eigenstate as well. If there exists such a
boson, irrespective of its parity, it can be directly observed in the decay
$J/\psi \to \mu^- \mu^+ X$ where $X$ remains invisible. We show that by using
the angular distribution or the distribution of events in the square Dalitz
plot, along with two well defined dimensionless ratios, one can clearly
distinguish among the various spin-parity possibilities. This would constitute
an important probe of both the existence and the nature of this new physics
possibility.
\end{abstract}


\maketitle

\section{Introduction}

It is well known that the anomalous magnetic moment of muon, $a_\mu =
\big(g_\mu-2\big)/2$ where $g_\mu$ is the $g$-factor of muon, is sensitive to
contributions from various new physics possibilities via quantum loops (see
Ref.~\cite{Lindner:2016bgg} for a recent review on contributions from various
specific new physics possibilities on $a_\mu$). Therefore, a difference between
the experimental measurement of $a_\mu$ and its standard model (SM) prediction
is considered to be a good probe of new physics. The precise experimental
measurement of muon anomalous magnetic moment as reported by the E821 experiment
$\Big(a_\mu^\textrm{exp}\Big)$ at Brookhaven National Laboratory (BNL)
\cite{Bennett:2006fi} is found to be larger than the existing most precise
theoretical prediction for the same from the SM $\Big(a_\mu^\textrm{SM}\Big)$
\cite{Aoyama:2020ynm}:
\begin{equation}\label{eq:mamm-anomaly}
\Delta a_\mu \equiv a_\mu^\textrm{exp} - a_\mu^\textrm{SM} = 279\left(76\right)\times 10^{-11}.
\end{equation}
This $3.7\sigma$ discrepancy can be considered as a tantalizing hint of the
presence of some new physics. It is expected that this discrepancy might
increase in the near future once results from the E989 experiment at Fermilab
\cite{Grange:2015fou} are available which would improve the existing
experimental measurement by a factor of four. The J-PARC New g-2/EDM experiment
at KEK, Japan \cite{Iinuma:2011zz} is also going to probe this discrepancy with
more precision. If the discrepancy goes beyond the accepted discovery threshold
of $5\sigma$ it would only imply that there is definitively some new physics
contributing via the quantum loops. However, to pin point the nature of the new
physics it is pertinent that we directly probe the various new physics
possibilities in other experiments.

\begin{figure}[hbtp]
\centering%
\includegraphics[scale=0.8]{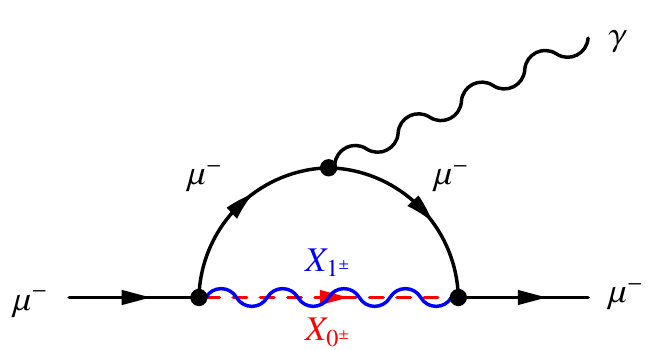}%
\caption{Contribution from the muon-philic boson $X$ to $\Delta a_\mu$.}
\label{fig:mamm}
\end{figure}

In one of the simplest new physics possibilities one introduces a new,
electrically neutral, spin-$0$ or spin-$1$ boson, say $X$, with exclusively
muon-philic interaction so as to avoid possible constraints from various other
experimental studies. The $X$ boson can have positive or negative parity, i.e.\
it could be a scalar, pseudo-scalar, vector or axial-vector particle, or it
could even not be a parity eigenstate. To denote these possibilities in a
unified manner let us write $X \equiv X_{s^\pm}$ where $s=0,1$ denotes the spin
and $\pm$ signify the parity. It is well known that for a scalar
$\left(X_{0^+}\right)$, pseudo-scalar $\left(X_{0^-}\right)$, vector
$\left(X_{1^+}\right)$ or axial-vector $\left(X_{1^-}\right)$ boson coupling
exclusively to muon via the following interaction Lagrangians,
\begin{subequations}\label{eq:Lagrangian}
\begin{align}
\mathscr{L}_{\mu}^{\textrm{S}} &= - g_{0+}^{} \, X_{0^+} \; \overline{\mu}\,\mu, \\%
\mathscr{L}_{\mu}^{\textrm{P}} &= - i \, g_{0-}^{} \, X_{0^-} \; \overline{\mu}\,\gamma^5\,\mu, \\%
\mathscr{L}_{\mu}^{\textrm{V}} &= - g_{1+}^{} \, \left(X_{1^+}\right)_\alpha \;
\overline{\mu}\,\gamma^\alpha\,\mu,\\%
\mathscr{L}_{\mu}^{\textrm{A}} &= - g_{1-}^{} \, \left(X_{1^-}\right)_\alpha \;
\overline{\mu}\,\gamma^\alpha\gamma^5\,\mu,%
\end{align} 
\end{subequations}
with mass $m_X \lesssim 2m_\mu$ where $m_\mu$ is the mass of muon (so that $X$
is invisible), we get the following contributions to $\Delta a_\mu$ at
leading-order \cite{Lindner:2016bgg} from Fig.~\ref{fig:mamm},
\begin{subequations}\label{eq:mamm}
\begin{align}
\Delta a_\mu^\textrm{S} &= \frac{g_{0+}^2}{8\pi^2} \int_{0}^{1} \frac{m_\mu^2 \,
\big(1-z\big) \, \big(1-z^2\big)}{m_\mu^2 \, \big(1-z\big)^2 + m_X^2 \, z} \,
\mathrm{d}z,\label{eq:mamm-S}\\%
\Delta a_\mu^\textrm{P} &= - \; \frac{g_{0-}^2}{8\pi^2} \int_{0}^{1} \frac{m_\mu^2
\, \big(1-z\big)^3}{m_\mu^2 \, \big(1-z\big)^2 + m_X^2 \, z}\,\mathrm{d}z,\label{eq:mamm-P}\\%
\Delta a_\mu^\textrm{V} &= \frac{g_{1+}^2}{8\pi^2} \int_{0}^{1} \frac{2 \,
m_\mu^2 \, z \, \big(1-z\big)^2}{m_\mu^2 \, \big(1-z\big)^2 + m_X^2 \,
z}\,\mathrm{d}z,\label{eq:mamm-V}\\%
\Delta a_\mu^\textrm{A} &= - \; \frac{g_{1-}^2}{8\pi^2} \int_{0}^{1} \frac{2 m_\mu^2 
\big(1-z\big) \big(m_X^2 z \left(3+z\right) + 2
m_\mu^2 \left(1-z\right)^2 \big)}{m_X^2 \, \left(m_\mu^2 \, \big(1-z\big)^2 +
m_X^2 \, z\right)}\,\mathrm{d}z,\label{eq:mamm-A}%
\end{align}
\end{subequations}
where $g_{s\pm}$ are the various coupling constants introduced in
Eq.~\eqref{eq:Lagrangian}. It is clear that when $X$ is not an eigenstate of
parity, it simply implies that we could have both scalar and pseudo-scalar
couplings for spin-$0$ case and both vector and axial-vector couplings for
spin-$1$ case. In such cases, one would have to add the corresponding
contributions to $\Delta a_\mu$ to explain the observed anomaly in Eq.~\eqref{eq:mamm-anomaly}.

There exist possible UV complete models that can provide interactions as
required in our Eq.~\eqref{eq:Lagrangian}, see Ref.~\cite{Lindner:2016bgg}.
Inevitably, these UV complete models introduce additional interactions which put
constraint on the relevant parameter space from other processes. For the scalar
case, one example is the type-$X$ two Higgs doublet model \cite{Barger:2009me,
Craig:2012vn, Craig:2013hca}. As was recently shown in \cite{Liu:2020qgx}, a
light scalar can accommodate both muon $g-2$ anomaly as well as the KOTO anomaly
in $K_L \to \pi^0 \nu \overline{\nu}$ decay \cite{Ahn:2018mvc, Shinohara:2019}.
Similarly, for the vector case, an example of UV complete model could be the
$U(1)_{L_\mu -L_\tau}$ model which conserves the difference between muon and tau
lepton number, in addition to being anomaly free \cite{He:1990pn, He:1991qd,
Foot:1994vd}. In a recent study \cite{Ban:2020uii} it was shown that such a
model could accommodate both muon $g-2$ anomaly and the anomaly reported by the
Atomki experiment in ${}^8\textrm{Be}^*$ \cite{Krasznahorkay:2015iga,
Krasznahorkay:2019lyl}. The mass of the new scalar or vector particles is in the
MeV region which is also the region of interest to us. However, in this paper we
are not concerned with the UV completion of the simplistic muon-philic model
that we have considered. Instead we will focus on how the existence of the $X$
boson can be studied and how its spin-parity can be determined. This would offer
a direct probe of the nature of the new physics under our consideration.

It is clear from Eq.~\eqref{eq:mamm} that the pseudo-scalar and axial-vector
contributions have opposite sign compared with contributions from scalar and
vector cases. This would imply that pseudo-scalar and axial-vector contributions
would push the theoretical value farther from the experimental value, and can
not explain the observed anomaly of Eq.~\eqref{eq:mamm-anomaly} when considered
separately. Therefore, purely from the point of view of muon anomalous magnetic
moment one usually considers only the scalar and vector new physics scenarios.
In a recent work of one of the authors \cite{Cvetic:2020vkk} it was explored how
the decay $J/\psi \to \mu^-\mu^+ X$ can indeed be used to probe this sort of new
physics in a conclusive and independent manner at the BES~III experiment. This
decay mode can be compared to the analogous process $e^+e^- \to \mu^- \mu^+ X$
proposed in context of Belle~II \cite{Jho:2019cxq} which has a large number of
competing background processes that must be carefully considered. The $X$ boson
in the final state is electrically neutral and does not decay (in the simplest
muon-philic scenario as exemplified by Eq.~\eqref{eq:Lagrangian} above) leaving
no track in the detector. As explained in detail in \cite{Cvetic:2020vkk} the
most dominant SM background for the decay $J/\psi \to \mu^-\mu^+ X$ arises from
final state radiation which can be readily handled experimentally by employing a
cut on the missing energy distribution. The feasibility of observing the decay
process is exemplified by the fact that around $30$ to $300$ events for the
scalar case and about $300$ to $2000$ events for the vector case are expected at
BES~III after the aforementioned missing energy cut is applied
\cite{Cvetic:2020vkk}.

It is important to note that our paper differs significantly from Ref.~\cite{Cvetic:2020vkk} by the facts that 
\begin{inparaenum}[(1)]
\item we consider all the four spin-parity possibilities of $X$ as well as the
possibility of $X$ not being a parity eigenstate, instead of only scalar and
vector cases as in \cite{Cvetic:2020vkk}, %
\item we show that for the case of $X$ being not a parity eigenstate the
parameter space allowed by muon $g-2$ is widened enhancing the branching ratios
for the decay $J/\psi \to \mu^- \mu^+ X$ making their experimental observation
more feasible, and %
\item we provide a completely different methodology to ascertain the various
spin-parity possibilities of $X$ using the difference in distribution patterns
in conventional Dalitz plot, angular distribution, square Dalitz plot as well as
by using two experimentally accessible dimensionless ratios.
\end{inparaenum}

Our paper is organized as follows. In Sec.~\ref{sec:allowed-parameter-space} we
find out the region of parameter space allowed by the existing anomaly in muon
anomalous magnetic moment, considering the simple muon-philic interactions of
Eq.~\eqref{eq:Lagrangian}. This is followed by an analytical analysis of the
decay $J/\psi \to \mu^- \mu^+ X$ for all the various spin-parity possibilities
of $X$ in Sec.~\ref{sec:analytical-study}. Along the way, in
Sec.~\ref{sec:numerical-study}, we numerically show the differences in patterns
of distribution of events in different kinds of Dalitz plots and angular
distributions for the various spin-parity possibilities, and discuss how they
can be used in conjunction with two dimensionless experimentally measurable
ratios to distinguish the various spin-parity possibilities. Finally we conclude
in Sec.~\ref{sec:conclusion} highlighting the main ideas and the future prospect
of our study.

\section{Constraints on pure muon-philic interactions from muon anomalous magnetic moment}\label{sec:allowed-parameter-space}

\begin{figure}[hbtp]
\centering%
\includegraphics[width=\linewidth, keepaspectratio]{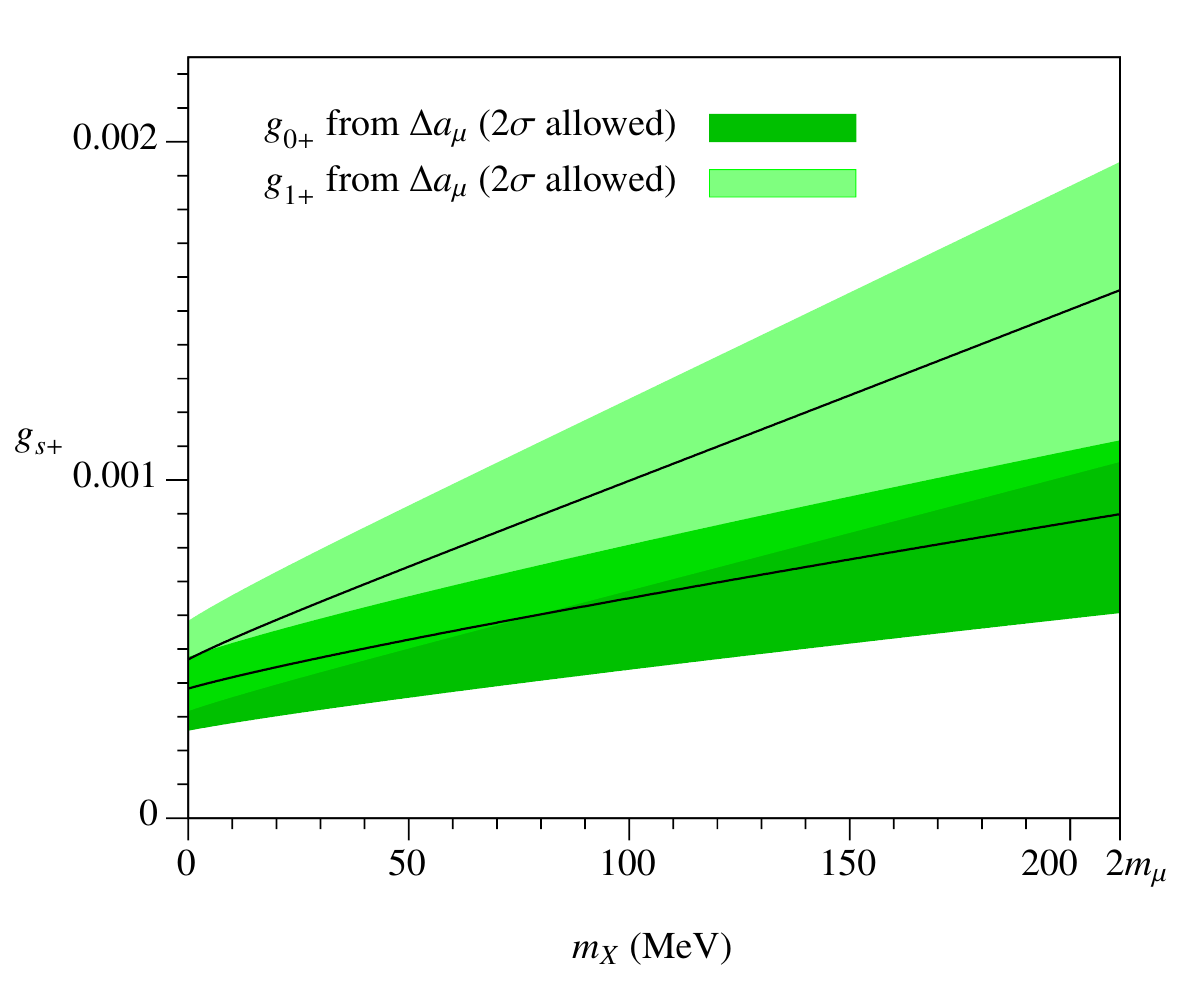}%
\caption{The coupling constants $g_{0+}^{}$ and $g_{1+}^{}$ as allowed by
$\Delta a_\mu$ at $2\sigma$ level for $m_X \lesssim 2m_\mu$. Here $X$ is
considered to be a parity eigenstate, either a scalar or vector.}%
\label{fig:SV-region}
\end{figure}

\begin{figure}[hbtp]
\centering%
\subfloat[A part of the allowed parameter space for $g_{0+}^{}$ and
$g_{0-}^{}$.\label{fig:SP}]{\includegraphics[width=\linewidth,
keepaspectratio]{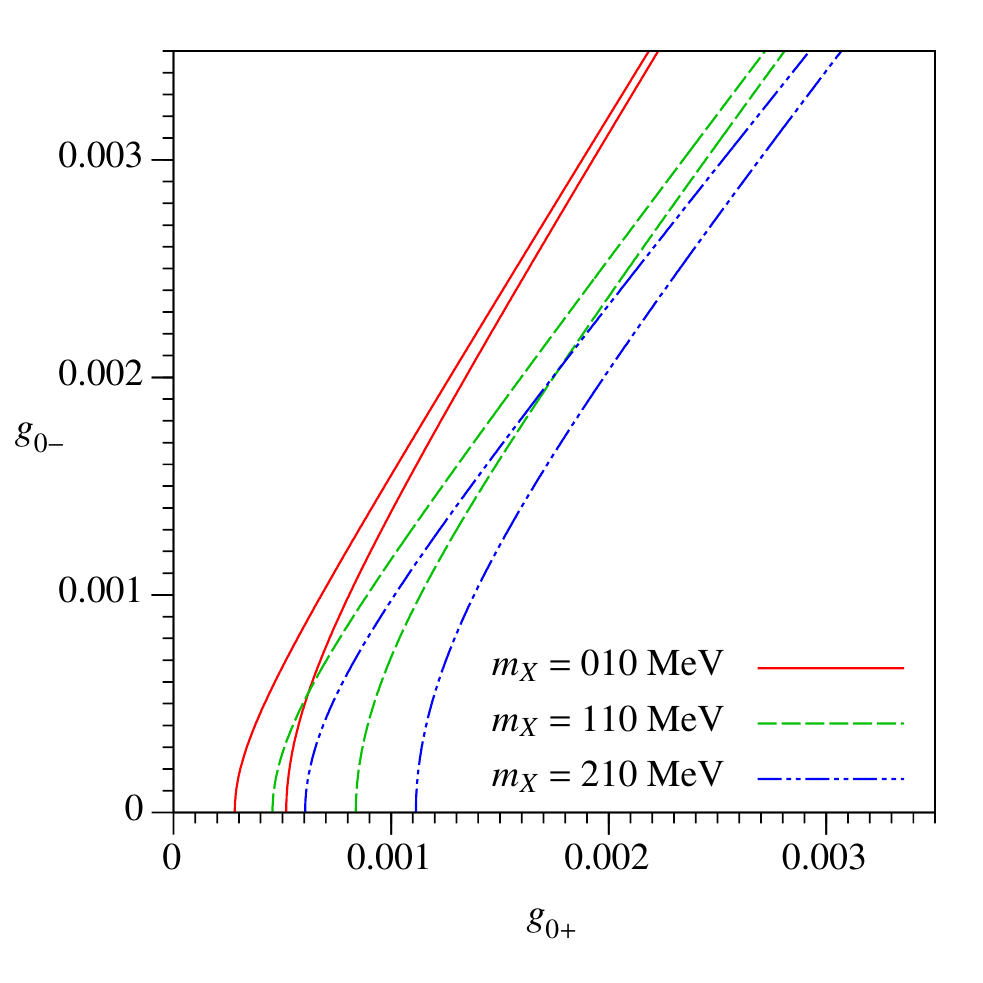}}\\%
\subfloat[A part of the allowed parameter space for $g_{1+}^{}$ and
$g_{1-}^{}$.\label{fig:VA}]{\includegraphics[width=\linewidth,
keepaspectratio]{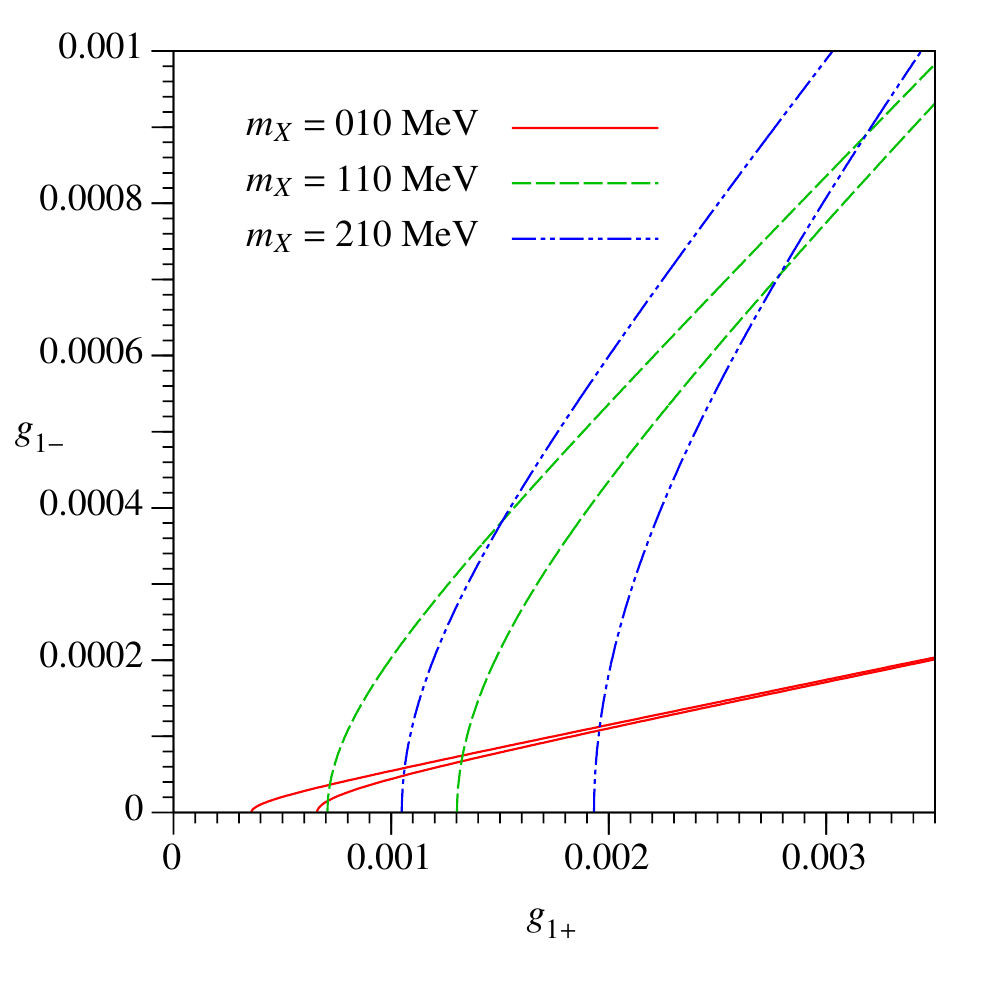}}%
\caption{The region of parameter space allowed by $\Delta a_\mu$ at $2\sigma$
level when $X$ is not a parity eigenstate, shown for a few chosen values of $m_X
\lesssim 2m_\mu$. The allowed region for the specific $m_X$ value is the region
bounded by the corresponding colored and dashed lines.}%
\label{fig:SPVA-region}
\end{figure}

By ascribing the anomaly in muon anomalous magnetic moment to muon-philic
interactions alone we can readily find out the allowed range of values for the
coupling constants $g_{s\pm}^{}$ where $s=0,1$. Allowing the current discrepancy
in Eq.~\eqref{eq:mamm-anomaly} at $2\sigma$ level and considering
Eq.~\eqref{eq:mamm} and taking the $X$ boson as a particle with definite parity,
the scalar coupling constant $g_{0+}^{}$ and the vector coupling constant
$g_{1+}^{}$ are allowed to vary in the green colored bands of
Fig.~\ref{fig:SV-region}.

If, instead, we relax the condition that the $X$ boson must be a parity
eigenstate, the coupling constants $g_{0+}^{}$ and $g_{0-}^{}$ are together
allowed for spin-$0$ case, and for spin-$1$ case both $g_{1+}^{}$ and
$g_{1-}^{}$ are allowed. Because of the fact that we have two parameters to fit
one data, the allowed region naturally widens, as can be seen from
Fig.~\ref{fig:SPVA-region}. Interestingly, it is the negative sign in front of
Eqs.~\eqref{eq:mamm-P} and \eqref{eq:mamm-A} which enable this widening in
parameter space. In Fig.~\ref{fig:SPVA-region} the region allowed by $\Delta
a_\mu$ at $2\sigma$ level is shown for the specific $m_X$ values of $10$~MeV,
$110$~MeV and $210$~MeV as the region bounded by the corresponding colored and
dashed lines. It is interesting that one can consider much larger values than
what is allowed if $X$ were to be a parity eigenstate. It is also clear from
Fig.~\ref{fig:SPVA-region} that the special cases of $g_{0+} = g_{0-}$ are also
allowed for most of the $m_X$ values.

Since the branching ratio of any process that would probe $X$ would be, in
general and at least, proportional to square of the coupling constant, a larger
value of the coupling constant would imply bigger branching ratio. Therefore,
the scenario where $X$ is not a parity eigenstate could be probed better in
experiment. Nevertheless, as we have seen from Figs.~\ref{fig:SV-region} and
\ref{fig:SPVA-region} the study of muon anomalous magnetic moment alone can not
decipher the nature of the new physics. In order to find out the nature of new
physics we need to supplement exploration of $\Delta a_\mu$ with some other
study, such as the decay $J/\psi \to \mu^-\mu^+ X$. By analyzing the Dalitz plot
distributions of $J/\psi \to \mu^-\mu^+ X$ or its angular distribution in the
center-of-momentum frame of $\mu^-\mu^+$ one should, in principle, distinguish
between the various new physics possibilities. In the following we develop this methodology in detail.

\section{Study of the decay \texorpdfstring{$\boldsymbol{J/\psi \to \mu^-\mu^+ X}$}{J/psi --> mu- mu+ X} with all spin-parity possibilities of \texorpdfstring{$\boldsymbol{X}$}{X}}\label{sec:analytical-study}

\begin{figure}[hbtp!]
\centering%
\includegraphics[scale=0.8]{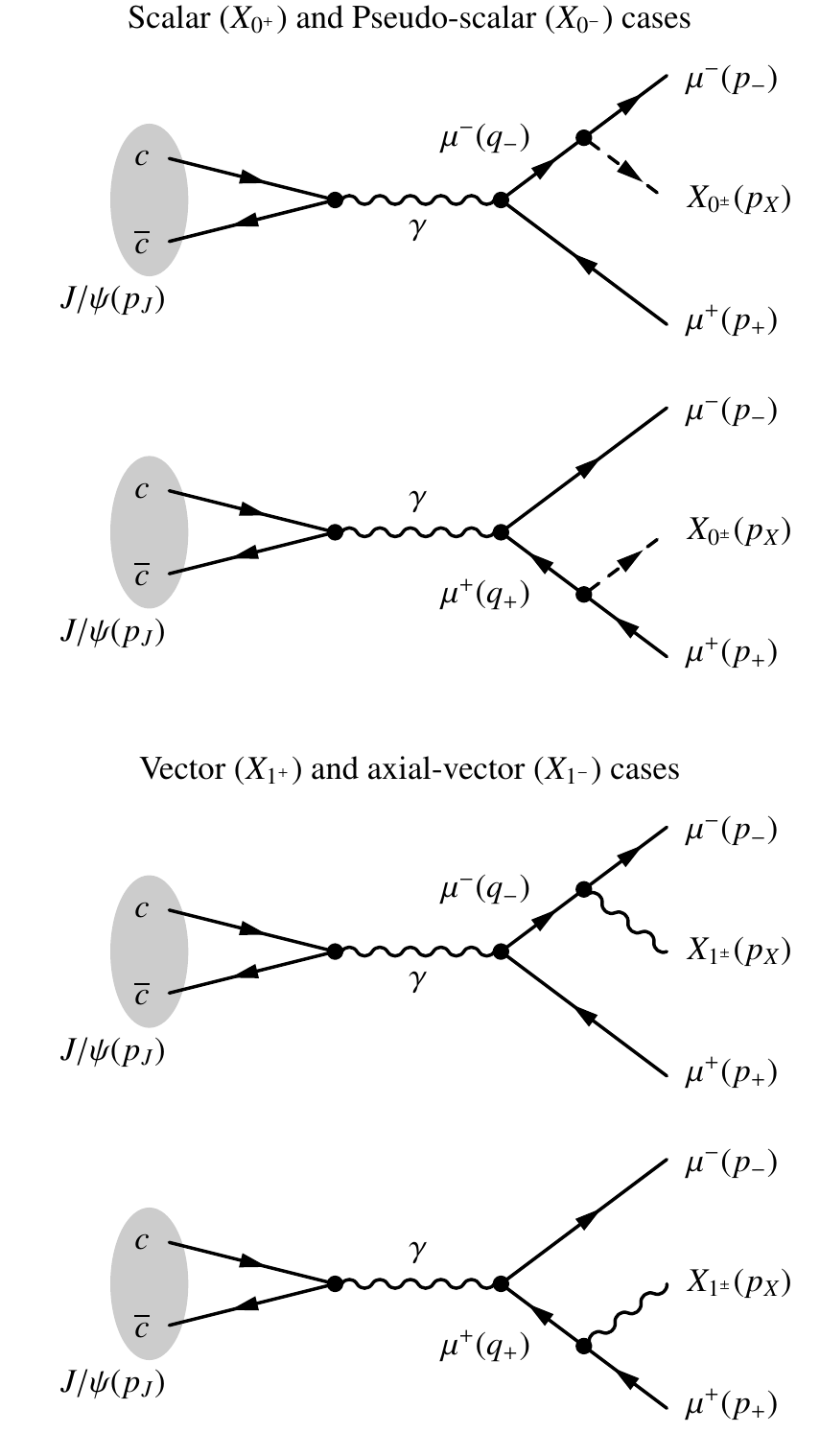}
\caption{Feynman diagrams for the decay $J/\psi \to \mu^- \mu^+ X$. The notation
for $4$-momenta of all the particles is also shown here.}%
\label{fig:Feynman-diagrams}
\end{figure}

The decay $J/\psi \to \mu^-\mu^+ X_{s^\pm}$ takes place via the Feynman diagrams
as shown in Fig.~\ref{fig:Feynman-diagrams}. The electrically neutral $X$ boson
can arise from either of the muon legs, and considering only the muon-philic
interactions of Eq.~\eqref{eq:mamm} it is stable and remains invisible in the
mass regime $m_X \lesssim 2m_\mu$ which is the region of our interest.
Considering the $X$ boson to be a parity eigenstate, it was shown in
Ref.~\cite{Cvetic:2020vkk} that this decay can be studied at BES~III. As noted
in the previous section, from Fig.~\ref{fig:SPVA-region} we find that when $X$
is not a parity eigenstate the coupling constants can have much larger values
and, therefore, it is evident that this possibility is experimentally
interesting.

Since we are considering a three-body decay, the allowed phase-space is fully
described by only two variables, such as two energies, or two invariant
mass-squares, or one invariant mass square and an angle. Below we consider a few of these pairs of variables and find out the expressions for corresponding distributions. 

\subsection{The normal Dalitz plot distribution in terms of two invariant mass-squares}

Denoting the $4$-momenta of $J/\psi$, $\mu^-$, $\mu^+$ and $X$ by $p_J$, $p_-$,
$p_+$ and $p_X$ respectively, let us define the following three invariant
mass-squares $s$, $t$ and $u$,
\begin{subequations}\label{eq:inv-mass-squares}
\begin{align}
s &= \left(p_+ + p_-\right)^2 = \left(p_J - p_X\right)^2,\label{eq:s}\\%
t &= \left(p_+ + p_X\right)^2 = \left(p_J - p_-\right)^2,\label{eq:t}\\%
u &= \left(p_- + p_X\right)^2 = \left(p_J - p_+\right)^2.\label{eq:u}
\end{align}
\end{subequations}
The three invariant mass-squares are not independent since $s+t+u = m_J^2 +
m_X^2 + 2 m_\mu^2$, where $m_J$ is the mass of $J/\psi$. Taking the invariant
mass-squares $t$ and $u$ as the two parameters, the differential decay rate for
the decay $J/\psi \to \mu^-\mu^+ X_{s^\pm}$ can be expressed by
\begin{equation}\label{eq:diff-decay-rate-tu}
\frac{\mathrm{d}^2 \Gamma_{s^\pm}}{\mathrm{d}t \; \mathrm{d}u} \equiv \frac{\mathrm{d}^2 \Gamma \left(J/\psi \to \mu^-\mu^+
X_{s^\pm}\right)}{\mathrm{d}t \; \mathrm{d}u} = \frac{\alpha^2\, g_{s\pm}^2 \,
f_J^2}{27\,\pi\, m_J^5 \, Y} \modulus{A_{s^\pm}}^2,
\end{equation}
where $\alpha$ is the fine-structure constant, $f_J = 0.407$~GeV \cite{Donald:2012ga} is the decay
constant of the $J/\psi$, $Y=\left(t-m_\mu^2\right)^2 \left(u-m_\mu^2\right)^2$, and
$\modulus{A_{s^\pm}}^2$ are some function of $t$, $u$ and the masses
involved, as given in Appendix~\ref{app:AspmSq}. 

All differences among the scalar, pseudo-scalar, vector and axial-vector
possibilities are present in the expressions for $\modulus{A_{s^\pm}}^2$ as
shown in Eq.~\eqref{eq:ASq}. The distribution expressed in
Eq.~\eqref{eq:diff-decay-rate-tu} (which is the normal Dalitz plot distribution)
is unique in the sense that the invariant mass-squares $t$ and $u$ do not depend
on the frame of reference. 

This must be noted that when $X$ is not a parity eigenstate, the differential
decay rate is easily obtained, analytically, by simply adding the two
contributions one would have if $X$ had definite parity, i.e.\
\begin{equation}\label{eq:noneigenstate-diff-decay-rate}
\frac{\mathrm{d}^2 \Gamma_{s}}{\mathrm{d}t
\; \mathrm{d}u} \equiv \frac{\mathrm{d}^2 \Gamma \left(J/\psi \to \mu^-\mu^+
X_{s}\right)}{\mathrm{d}t \; \mathrm{d}u} = \frac{\mathrm{d}^2 \Gamma_{s^+}}{\mathrm{d}t \; \mathrm{d}u} + \frac{\mathrm{d}^2 \Gamma_{s^-}}{\mathrm{d}t \; \mathrm{d}u}.
\end{equation}
From Eq.~\eqref{eq:noneigenstate-diff-decay-rate} it is clear that, in general,
$\Gamma \left(J/\psi \to \mu^-\mu^+ X_{s}\right) > \Gamma \left(J/\psi \to
\mu^-\mu^+ X_{s^+}\right)$. The decay rate $\Gamma \left(J/\psi \to \mu^-\mu^+
X_{s}\right)$ also gets further enhanced due to the fact that when $X$ is not a
parity eigenstate the allowed parameter space has a much wider spread as seen
from Fig.~\ref{fig:SPVA-region} vis-\`{a}-vis Fig.~\ref{fig:SV-region}. For
example, the value of $g_{0+}$ could be about $5$ times larger when $X$ is not a
parity eigenstate than when $X$ is purely scalar, and thus one can expect that
the number of $J/\psi \to \mu^-\mu^+ X_{0}$ events for the generic spin-$0$ case
would be larger than $25$ times the number of $J/\psi \to \mu^-\mu^+ X_{0^+}$
events for the scalar case. Similar arguments also hold for the spin-$1$ case.

Instead of the $t$ vs.\ $u$ Dalitz plot considered above, one could also think
of another normal Dalitz plot using $E_-$ and $E_+$, where $E_\pm $ denotes the
energy of $\mu^\pm$ in the rest frame of $J/\psi$. These two Dalitz plot
distributions are related to one another by
\begin{equation}\label{eq:diff-decay-rate-Epm}
\frac{\mathrm{d}^2\Gamma_{s^\pm}}{\mathrm{d}E_- \; \mathrm{d}E_+} =
4\,m_J^2\,\frac{\mathrm{d}^2\Gamma_{s^\pm}}{\mathrm{d}t \; \mathrm{d}u},
\end{equation}
arising from the simple observation that in the rest frame of $J/\psi$ we have
\begin{subequations}\label{eq:tu}
\begin{align}
t &= m_J^2 + m_\mu^2 - 2 m_J E_-,\label{eq:Em}\\%
u &= m_J^2 + m_\mu^2 - 2 m_J E_+.\label{eq:Ep}
\end{align}
\end{subequations}
For our analysis either of these normal Dalitz plots would serve the same
purpose. However, for our numerical study in Sec.~\ref{sec:numerical-study} we
have chosen the $t$ vs.\ $u$ Dalitz plot due to its Lorentz invariant nature.

\subsection{The angular distribution in the center-of-momentum frame of the muon pair}

\begin{figure}[hbtp]
\centering%
\includegraphics[scale=0.8]{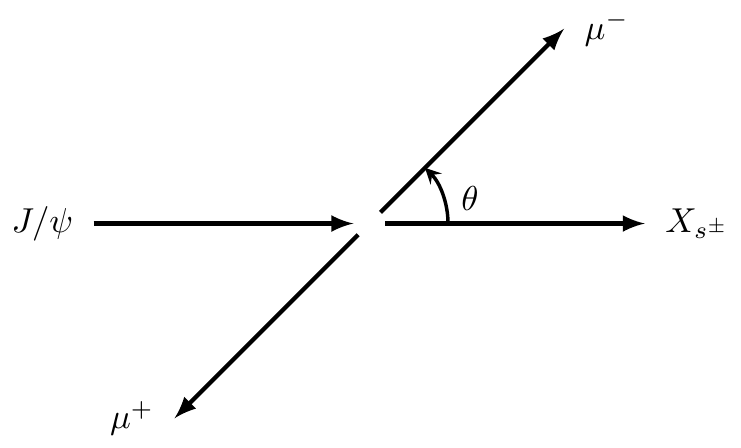}%
\caption{The center-of-momentum frame of the muon pair, equivalently the Gottfried-Jackson frame.}
\label{fig:GJ-frame}
\end{figure}

To arrive at another useful distribution we need to consider the
Gottfried-Jackson frame of reference, which is essentially the
center-of-momentum frame of the muon pair. Let the angle subtended by the
direction of flight of $\mu^-$ in this frame with respect to the direction of
flight of $J/\psi$ (equivalently that of $X_{s^\pm}$ as well) be $\theta$, see
Fig.~\ref{fig:GJ-frame}. In this frame of reference we find that,
\begin{subequations}\label{eq:tu-GJ}
\begin{align}
t &= a + b\cos\theta,\label{eq:t-GJ}\\%
u &= a - b \cos\theta,\label{eq:u-GJ}
\end{align}
\end{subequations}
where
\begin{subequations}\label{eq:ab}
\begin{align}
a &= \frac{1}{2} \left(m_J^2 + m_X^2 + 2 m_\mu^2 - s\right),\label{eq:a}\\%
b &= \frac{1}{2} \sqrt{1-\frac{4m_\mu^2}{s}} \sqrt{\lambda\left(s,m_J^2,m_X^2\right)},\label{eq:b}
\end{align}
\end{subequations}
with the K\"{a}ll\'{e}n function being given by $$\lambda \left(x,y,z\right) =
x^2 + y^2 + z^2 - 2 \left(xy+yz+zx\right).$$

Using Eqs.~\eqref{eq:tu-GJ} and
\eqref{eq:ab} we can make a change of variables and write down the following
expression for the differential decay rate in the Gottfried-Jackson frame in
terms of $s$ and $\cos\theta$,
\begin{equation}\label{eq:diff-decay-rate-scth}
\frac{\mathrm{d}^2\Gamma_{s^\pm}}{\mathrm{d}s \; \mathrm{d}\cos\theta} =
\frac{2\,m_J\,b\,\sqrt{s}}{m_J^2+s-m_X^2} \frac{\mathrm{d}^2 \Gamma_{s^\pm}}{\mathrm{d}t \; \mathrm{d}u}.
\end{equation}
Note that while going from Eq.~\eqref{eq:diff-decay-rate-tu} to
\eqref{eq:diff-decay-rate-scth} the effect of time dilation has been taken into
consideration, and the frame dependency of Eq.~\eqref{eq:diff-decay-rate-scth}
essentially stems from terms proportional to various powers of $\cos\theta$
after substituting $t$ and $u$ using Eq.~\eqref{eq:tu-GJ}.

\subsection{The square Dalitz plot distribution}

Another alternative distribution, favored by experimentalists, is the one called
square Dalitz plot. The idea of square Dalitz plot first appeared in a BaBar
paper \cite{Aubert:2005sk}, and recently it has been used in a LHCb study
\cite{Aaij:2014baa} as well. For our case, the expression for distribution of
events in the square Dalitz plot can be obtained from
Eq.~\eqref{eq:diff-decay-rate-scth} by making the following change of variables,
\begin{equation}
m' = \frac{1}{\pi} \arccos\left(2\frac{\sqrt{s}-2m_\mu}{m_J-m_X-2m_\mu}-1\right),\qquad
\theta' = \frac{1}{\pi} \theta,
\end{equation}
such that both $m'$ and $\theta'$ vary between $0$ and $1$. It is easy to show that
\begin{equation}\label{eq:diff-decay-rate-sdp}
\frac{\mathrm{d}^2\Gamma_{s^\pm}}{\mathrm{d}m' \; \mathrm{d}\theta'} = Z
\frac{\mathrm{d}^2\Gamma_{s^\pm}}{\mathrm{d}s \; \mathrm{d}\cos\theta},
\end{equation}
where
\begin{align}
Z &= \frac{\pi^2}{2} \sin\left(\pi\,m'\right) \; \sin\left(\pi\,\theta'\right)\;\Big(m_J - m_X-2m_\mu\Big)\nonumber\\%
&\quad \times \left(\Big(m_J-m_X-2m_\mu\Big) \, \Big(1+\cos\left(\pi\,m'\right)\Big) + 4 m_\mu \right).
\end{align}
This square Dalitz plot provides a clearer view of the peaks in the differential
decay rate as compared to the normal Dalitz plot and as we will see in
Sec.~\ref{sec:numerical-study} they play a vital role in our study.

\subsection{Information in the angular distribution}
Before we study the patterns in the various distributions mentioned above, it is
helpful to do a detailed analysis of the angular distribution,
Eq.~\eqref{eq:diff-decay-rate-scth}, in particular. We find that the angular
distribution for all the spin-parity possibilities can be described in a unified
manner by using the orthogonal Legendre polynomials in the following
manner,
\begin{equation}\label{eq:angular-dist}
\frac{\mathrm{d}^2\Gamma_{s^\pm}}{\mathrm{d}s\;\mathrm{d}\cos\theta} =
\frac{g_{s\pm}^2\,C}{Y} \bigg(\mathcal{T}_{s\pm} +
\mathcal{U}_{s\pm}\,P_2\left(\cos\theta\right) +
\mathcal{V}_{s\pm}\,P_4\left(\cos\theta\right)\bigg),
\end{equation}
where 
\begin{equation}
C = \left(\frac{2\,\alpha^2\,f_J^2}{27\,\pi\,m_J^4}\right) 
\left(\frac{b\,\sqrt{s}}{m_J^2+s-m_X^2}\right),
\end{equation}
$P_2(x) = \tfrac{1}{2}\left(3x^2-1\right)$,
$P_4(x)=\tfrac{1}{8}\left(35x^4-30x^2+3\right)$ denote the Legendre polynomials
of order $2$ and $4$ respectively, and the angular coefficients
$\mathcal{T}_{s\pm}$, $\mathcal{U}_{s\pm}$ and $\mathcal{V}_{s\pm}$ are as given
in Appendix~\ref{app:angular-coefficients}.

Due to the orthogonal nature of the Legendre polynomials in
Eq.~\eqref{eq:angular-dist}, all the individual terms  $\mathcal{T}_{s\pm}$,
$\mathcal{U}_{s\pm}$ and $\mathcal{V}_{s\pm}$ explicitly shown in
Eq.~\eqref{eq:angular_coefficients} of Appendix~\ref{app:angular-coefficients}
can be obtained via the following integrations following the method of moments \cite{Gratrex:2015hna, Beaujean:2015xea},
\begin{subequations}\label{eq:integrations}
\begin{align}
\mathcal{T}_{s\pm} &= \frac{1}{2\,g_{s\pm}^2C} \int_{-1}^{1} Y\,
\frac{\mathrm{d}^2\Gamma_{s^\pm}}{\mathrm{d}s \; \mathrm{d}\cos\theta}\;
\mathrm{d}\cos\theta,\\%
\mathcal{U}_{s\pm} &= \frac{5}{2\,g_{s\pm}^2\,C} \int_{-1}^{1}
Y\,P_2\left(\cos\theta\right)\, \frac{\mathrm{d}^2\Gamma_{s^\pm}}{\mathrm{d}s \; \mathrm{d}\cos\theta}\;
\mathrm{d}\cos\theta,\\%
\mathcal{V}_{s\pm} &= \frac{9}{2\,g_{s\pm}^2\,C} \int_{-1}^{1}
Y\,P_4\left(\cos\theta\right)\, \frac{\mathrm{d}^2\Gamma_{s^\pm}}{\mathrm{d}s \; \mathrm{d}\cos\theta}\;
\mathrm{d}\cos\theta.%
\end{align}
\end{subequations}
Extraction of these angular coefficients by Eq.~\eqref{eq:integrations} help us to define the following two ratios,
\begin{equation}\label{eq:ratios}
\mathcal{R}_1^{s\pm} = \mathcal{U}_{s\pm}/\mathcal{T}_{s\pm},\qquad
\mathcal{R}_2^{s\pm} = \mathcal{V}_{s\pm}/\mathcal{T}_{s\pm},
\end{equation}
which are independent of the coupling constants $g_{s\pm}^{}$. From
Eqs.~\eqref{eq:VtermS} and \eqref{eq:VtermP} in
Appendix~\ref{app:angular-coefficients} it is clear that for spin-$0$ case (both
scalar and pseudo-scalar) $R_2^{0\pm}=0$ while from Eqs.~\eqref{eq:VtermV} and
\eqref{eq:VtermA} we get $R_2^{1\pm} \neq 0$ for spin-$1$ case. Therefore,
$R_2^{s\pm}$ can easily distinguish between spin-$0$ and spin-$1$ cases.
Nevertheless, the ratio $R_1^{s\pm}$ also has distinct features for the various
spin-parity possibilities. We shall numerically show in the next section that
these two ratios can indeed be used to distinguish the various spin-parity
possibilities.

It should be noted that when $X$ is not a parity eigenstate, the angular distribution would have the following form,
\begin{align}
\frac{\mathrm{d}^2\Gamma_{s}}{\mathrm{d}s\;\mathrm{d}\cos\theta}=
\frac{C}{Y}\Bigg(\mathcal{T}_{s} +
\mathcal{U}_{s}\,P_2\left(\cos\theta\right) +
\mathcal{V}_{s}\,P_4\left(\cos\theta\right)\Bigg),\label{eq:noneigenstate-angular-dist}
\end{align}
where
\begin{subequations}
\begin{align}
\mathcal{T}_{s} &= g_{s+}^2\,\mathcal{T}_{s+} + g_{s-}^2\,\mathcal{T}_{s-},\\%
\mathcal{U}_{s} &= g_{s+}^2\,\mathcal{U}_{s+} + g_{s-}^2\,\mathcal{U}_{s-},\\%
\mathcal{V}_{s} &= g_{s+}^2\,\mathcal{V}_{s+} + g_{s-}^2\,\mathcal{V}_{s-}.%
\end{align}
\end{subequations}
We can also define two ratios similar to the ones in Eq.~\eqref{eq:ratios},
\begin{equation}\label{eq:noneigenstate-ratios}
\mathcal{R}_1^s = \mathcal{U}_{s}/\mathcal{T}_{s}, \qquad%
\mathcal{R}_2^s = \mathcal{V}_{s}/\mathcal{T}_{s},
\end{equation}
which, however, do depend on the coupling constants $g_{s\pm}^{}$. Interestingly
when $X$ is not a parity eigenstate, we only need to know its spin, which can be
easily inferred from $\mathcal{R}_2^s$ alone, since for spin-$0$ case
$\mathcal{R}_2^0 = 0$ and for spin-$1$ case $\mathcal{R}_2^1 \neq 0$.

It should be noted that although the angular distribution of
Eq.~\eqref{eq:diff-decay-rate-scth} is defined in the Gottfried-Jackson frame,
the ratios as defined in Eqs.~\eqref{eq:ratios} and
\eqref{eq:noneigenstate-ratios} are valid in all frames of reference, since they
are function of the invariant mass-square $s$ alone.

It is possible to do the integration over $s$ and
define the following ratios as well,
\begin{equation}
\left\langle \mathcal{R}_{1}^{s\pm} \right\rangle = \frac{\int ds\, \mathcal{U}_{s\pm}}{\int ds\, \mathcal{T}_{s\pm}},\qquad%
\left\langle \mathcal{R}_{2}^{s\pm} \right\rangle = \frac{\int ds\, \mathcal{V}_{s\pm}}{\int ds\, \mathcal{T}_{s\pm}}.
\end{equation}
Once again, for the spin-$0$ case we expect $\left\langle \mathcal{R}_{2}^{0\pm}
\right\rangle=0$ and for the spin-$1$ case $\left\langle \mathcal{R}_{2}^{1\pm}
\right\rangle \neq 0$.

There is one easy prescription to numerically measure the ratios $\left\langle
\mathcal{R}_{1,2}^{s\pm} \right\rangle$. For this we can consider the $t$ vs.\
$u$ Dalitz plot, or the $s$ vs.\ $\cos\theta$ distribution of events. For every
event we calculate the values of $Y=\left(\left(a-m_\mu^2\right)^2 -
b^2\,\cos^2\theta\right)^2$, $P_2\left(\cos\theta\right)$ and
$P_4\left(\cos\theta\right)$. Then we sum over all the values of $Y$,
$Y\,P_2\left(\cos\theta\right)$ and $Y\,P_4\left(\cos\theta\right)$ to obtain
$\left\langle Y \right\rangle$, $\left\langle Y P_2\left(\cos\theta\right)
\right\rangle$ and $\left\langle Y P_4\left(\cos\theta\right) \right\rangle$
respectively. Finally we use the following expressions to measure the values of
$\left\langle \mathcal{R}_{1,2}^{s\pm} \right\rangle$,
\begin{equation}
\left\langle \mathcal{R}_{1}^{s\pm} \right\rangle = \frac{\left\langle Y\,
P_2\left(\cos\theta\right) \right\rangle}{\left\langle Y \right\rangle},\qquad%
\left\langle \mathcal{R}_{2}^{s\pm} \right\rangle = \frac{\left\langle Y\,
P_4\left(\cos\theta\right) \right\rangle}{\left\langle Y \right\rangle}.
\end{equation}
Within experimental accuracy, if $\left\langle \mathcal{R}_{2}^{s\pm}
\right\rangle = 0$, then we have a spin-$0$ $X$ boson, else the $X$ boson has
spin-$1$.

\section{Numerical study}\label{sec:numerical-study}

\subsection{Canonical branching ratios and expected number of events at BESIII}

\begin{figure}[hbtp]
\centering%
\includegraphics[width=\linewidth, keepaspectratio]{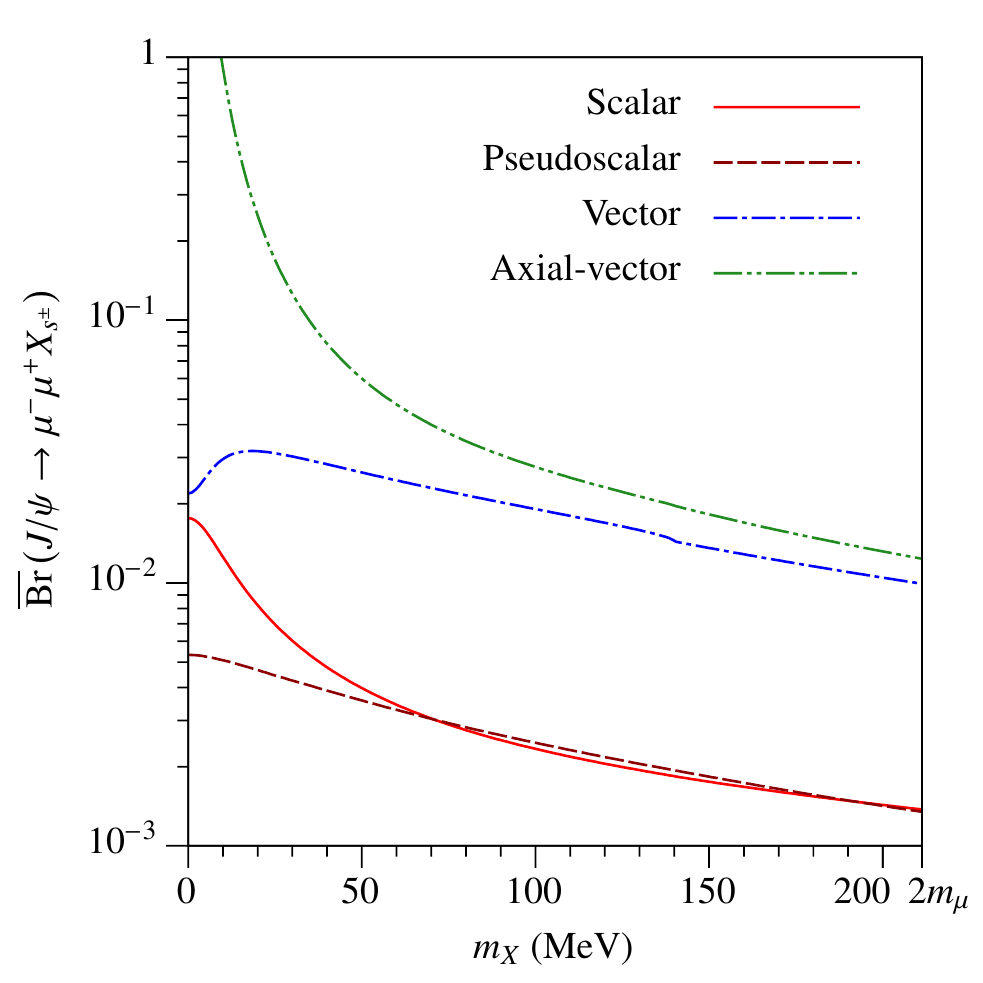}%
\caption{The canonical branching ratios, as defined in
Eq.~\eqref{eq:canonical-Br}, for various spin parity possibilities of $X$. The
evaluation includes a cut on the missing energy at $140$~MeV to eliminate the
significant SM background from final state radiation arising from $J/\psi \to
\mu^- \mu^+ \gamma$ events.}%
\label{fig:canonical-Br}
\end{figure}

Before we analyze the distribution patterns and ratios to distinguish the
various spin-parity possibilities of $X$, it would be helpful to know whether
such decays can indeed be probed experimentally. For this study we consider the
BESIII experiment as a natural choice because at BESIII a large number $(\sim
10^{11})$ of on-shell $J/\psi$ mesons will be produced at rest
\cite{BESIII:2019, Yuan:2019zfo, Ablikim:2019hff} which would help in inferring
the $4$-momentum of the invisible $X$ in a much easy and clean manner. Let us denote the
canonical branching ratios for the decays $J/\psi \to \mu^-\mu^+ X_{s^\pm}$ by $\overline{\textup{Br}}\left(J/\psi \to \mu^- \mu^+ X_{s^\pm}\right)$ and they are defined as the usual branching ratios sans the coupling constants i.e.\
\begin{equation}\label{eq:canonical-Br}
\overline{\textup{Br}}\left(J/\psi \to \mu^- \mu^+ X_{s^\pm}\right) = \frac{1}{g_{s\pm}^2} \frac{\Gamma\left(J/\psi \to \mu^- \mu^+ X_{s^\pm}\right)}{\Gamma_{J/\psi}},
\end{equation}
where $\Gamma_{J/\psi}$ is the experimentally measured total decay rate of
$J/\psi$. In Fig.~\ref{fig:canonical-Br} we have shown the variation of the
canonical branching ratios with the mass of $X$ for its various spin-parity
possibilities. The distinct trend of rise in the canonical branching ratio for
the axial-vector case in the limit $m_X \to 0$ is easily discernible from the
$m_X^{-2}$ factor in Eq.~\eqref{eq:ASq-A}. It must be noted that in evaluation
of the canonical branching ratios shown in Fig.~\ref{fig:canonical-Br} we have
included a cut on the missing energy at $E_\textrm{missing} = 140$~MeV as
suggested in Ref.~\cite{Cvetic:2020vkk} to completely eliminate the SM
background events arising from final state radiation in the form of $J/\psi \to
\mu^-\mu^+\gamma$, especially the soft photon ones since at BESIII photons with
energy less than $20$~MeV can not be detected at the detector
\cite{Ablikim:2009aa}. Details of the background study can be found in
Ref.~\cite{Cvetic:2020vkk}.

\begin{figure*}[hbtp]
\centering%
\includegraphics[width=0.475\linewidth]{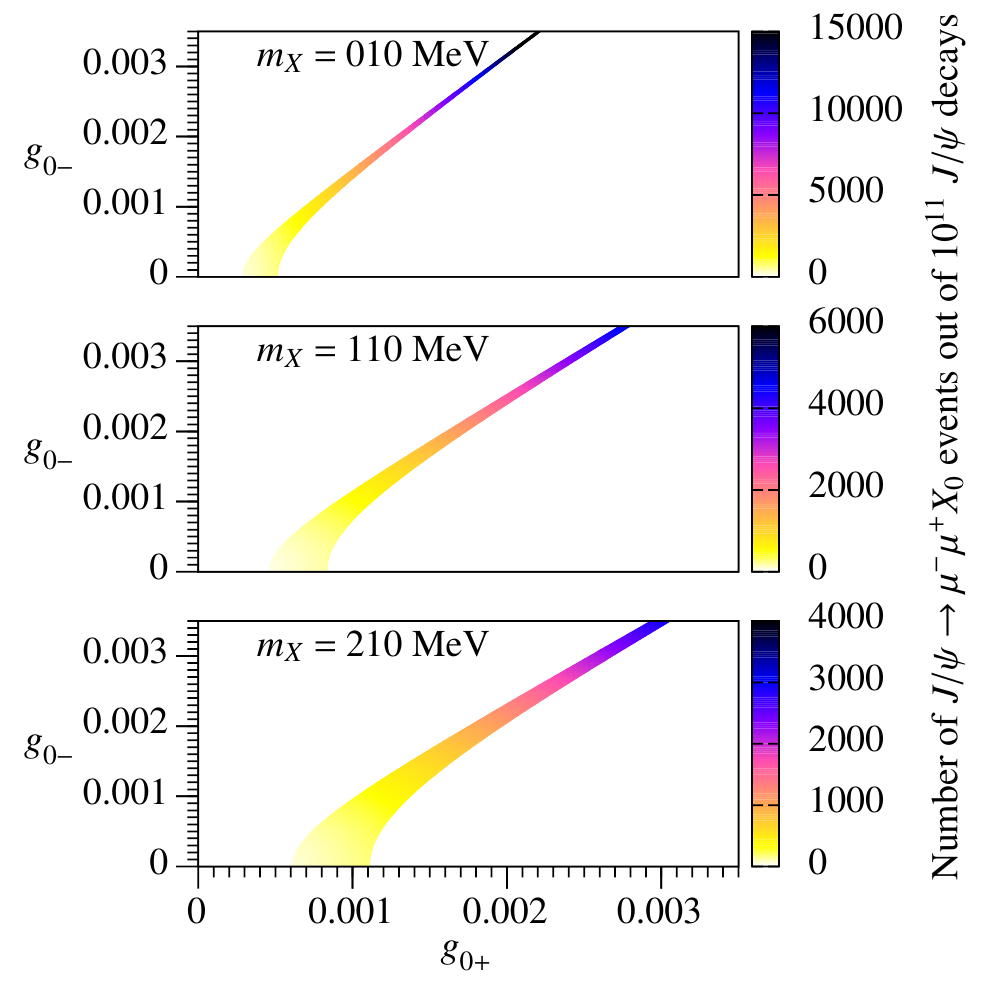}\hfill
\includegraphics[width=0.475\linewidth]{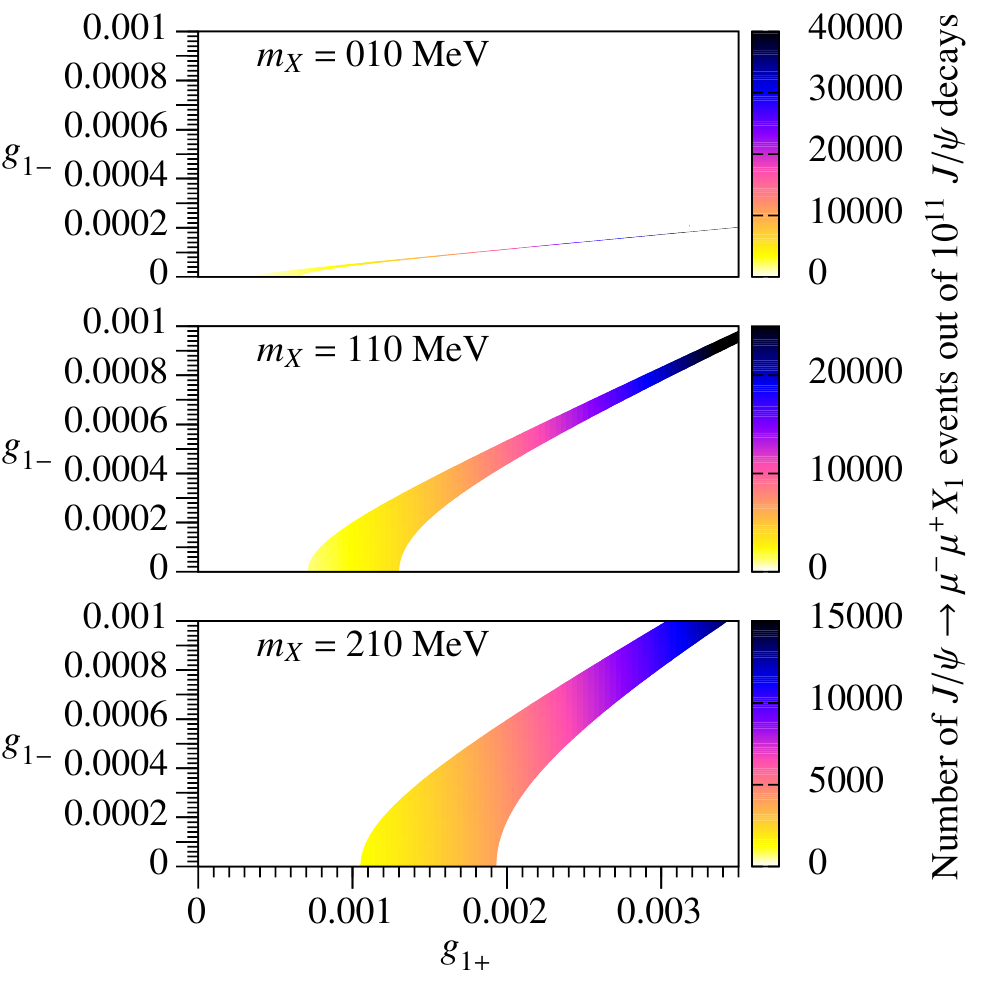}%
\caption{Expected number of $J/\psi \to \mu^-\mu^+ X_s$ events out of a total of
$10^{11}$ number of $J/\psi$ decays including a missing energy cut at $140$~MeV.}%
\label{fig:number-of-events}
\end{figure*}

The canonical branching ratios are helpful in predicting the number of signal
events, i.e.\ number of $J/\psi \to \mu^- \mu^+ X_{s}$ decays, one would
expect at BESIII if the muon-philic $X$ boson were to describe the muon $g-2$
anomaly. In Fig.~\ref{fig:number-of-events} we show the expected number of such
signal events across the parameter space allowed by anomalous magnetic moment of
muon (region shown before in Fig.~\ref{fig:SPVA-region}), for three
representative masses $m_X=10$~MeV, $110$~MeV and $210$~MeV. The high number of
events shown in Fig.~\ref{fig:number-of-events} are possible only when one
considers the $X$ boson to not be a parity eigenstate. This is due to the
widening of the parameter space allowed by the presence of additional coupling
constants $g_{0-}$ and $g_{1-}$ for the spin-$0$ and spin-$1$ cases
respectively.

It is nevertheless interesting and important to numerically study and find out
how to distinguish the individual spin-parity possibilities irrespective of the
muon anomalous magnetic moment consideration. For this we focus on a numerical
study of the various distributions and the ratios that were presented,
analytically, in the previous section.

\subsection{Patterns in distributions, ratios and discussion}

To study how the patterns of distribution differ for the different spin-parity
possibilities of $X$ as well as how these patterns evolve when we go from lower
values of $m_X$ to higher values of $m_X$, we once again consider the three
representative masses, $m_X=10$~MeV, $110$~MeV and $210$~MeV. We will analyze
the $t$ vs.\ $u$ Dalitz plot distribution of Eq.~\eqref{eq:diff-decay-rate-tu},
the $s$ vs.\ $\cos\theta$ angular distribution of
Eq.~\eqref{eq:diff-decay-rate-scth} and the $m'$ vs.\ $\theta'$ distribution
also known as the square Dalitz plot distribution as described by
Eq.~\eqref{eq:diff-decay-rate-sdp}. To make a meaningful  comparison across the
various spin-parity possibilities, we have normalized the distributions by their
maximum values which helps the patterns become more visually acute and easy to
grasp. Due to this normalization by the maximum value of the distribution, all
the distributions shown do not depend on the size of the coupling constants
$g_{s\pm}^{}$.

To put the distribution patterns into proper perspective we have also included
distribution of $1000$ simulated Monte Carlo events generated after applying a
cut on the missing energy at $140$~MeV as mentioned before.

\begin{turnpage}
\begin{figure*}[hbtp]
\includegraphics[height=0.9\textheight,width=0.33\linewidth,keepaspectratio]{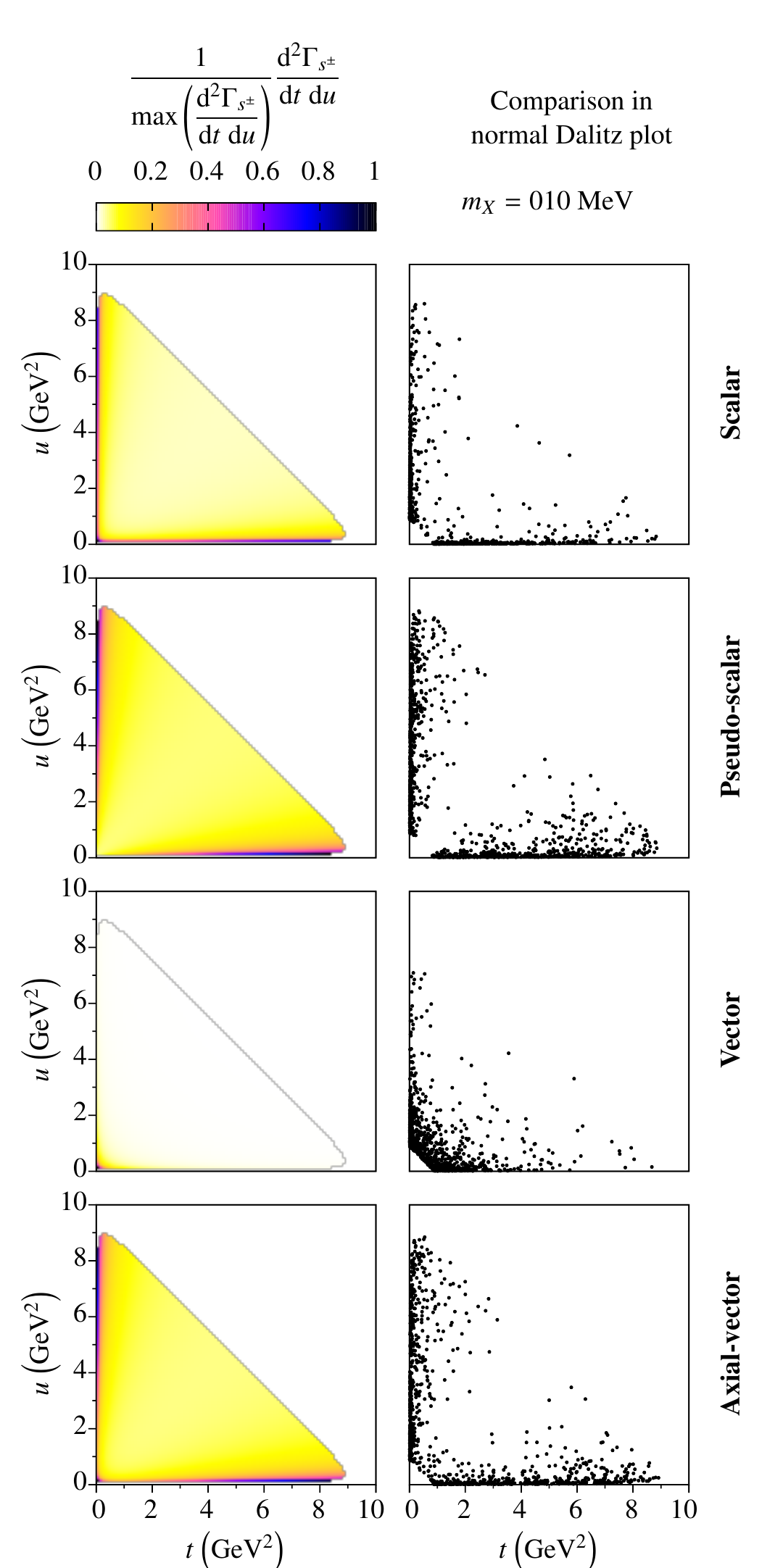}%
\includegraphics[height=0.9\textheight,width=0.33\linewidth,keepaspectratio]{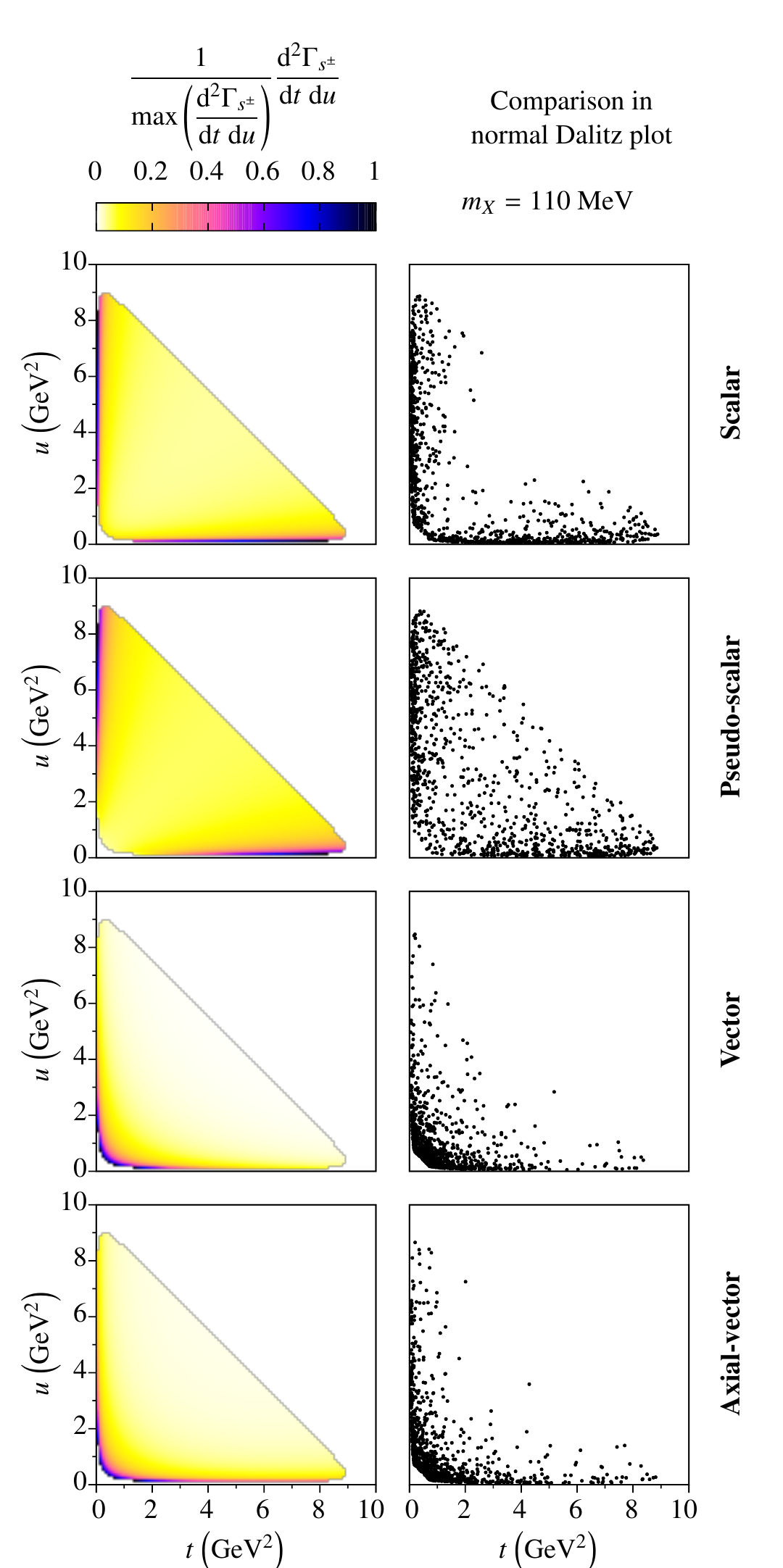}%
\includegraphics[height=0.9\textheight,width=0.33\linewidth,keepaspectratio]{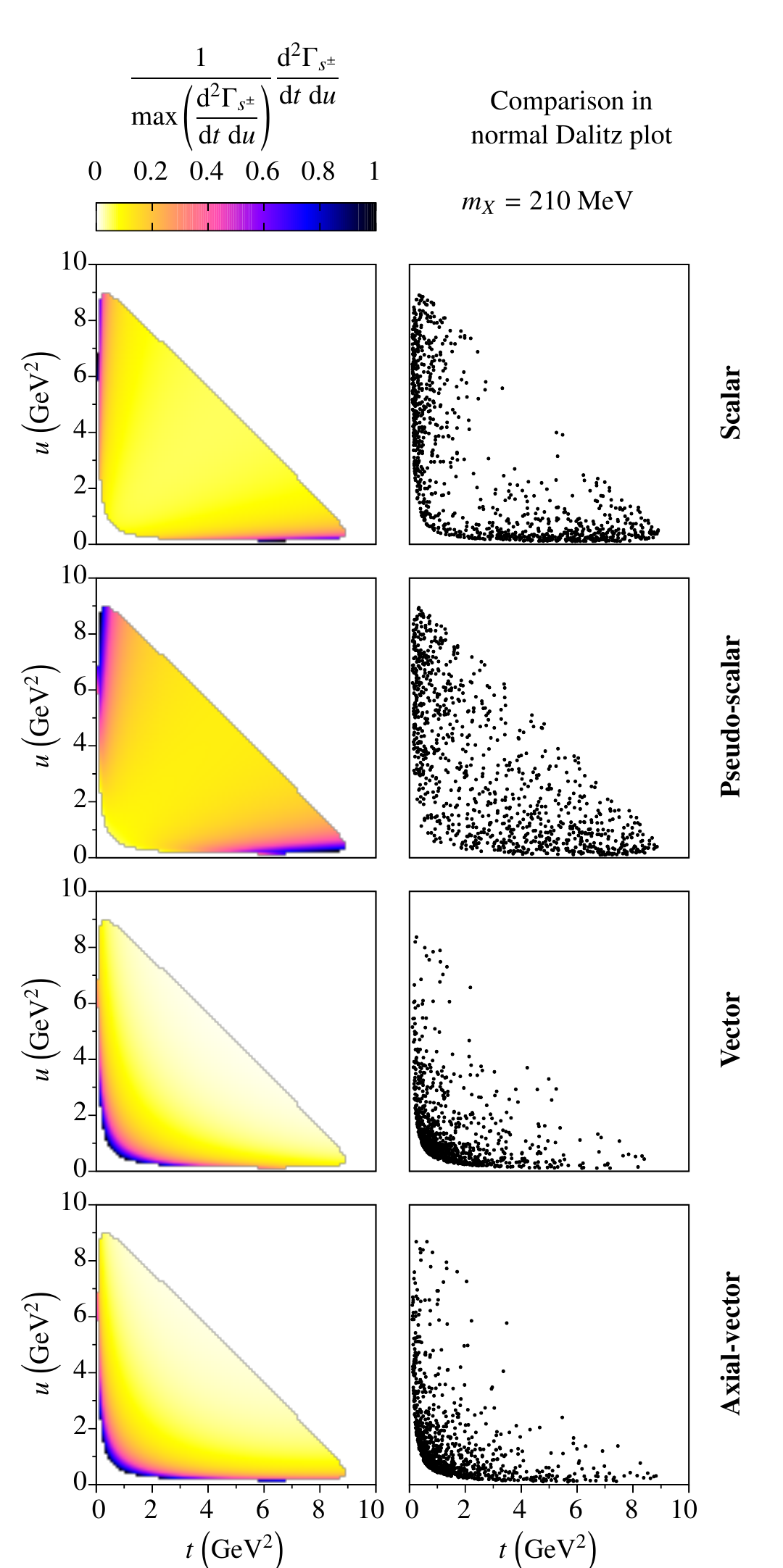}%
\caption{Expected patterns in normal $t$ vs.\ $u$ Dalitz plot distributions are
shown on the left panel of each of the subfigures, and on the right panel are
the distributions of $1000$ simulated events after applying a missing energy cut
at $140$~MeV.}%
\label{fig:Dalitz-distributions}
\end{figure*}
\end{turnpage}

\paragraph{\bf Normal Dalitz plot distribution:}
Let us first consider the $t$ vs.\ $u$ Dalitz plots, see
Fig.~\ref{fig:Dalitz-distributions}. It is clear that for $m_X=10$~MeV, all the
four distributions, corresponding to the four distinct spin-parity
possibilities, are extremely flat in most of the allowed region except at the
kinematic boundary. If we go to $m_X=110$~MeV, the distribution peaks are found
to be indeed close to the boundary, but are more prominent than before. Here we
can see that for spin-$0$ case there is no peak close to origin, while for
spin-$1$ case the peak sits nearer to the origin. Considering $m_X=210$~MeV we
find that the distribution still retains the characteristic locations of the
peaks as mentioned before, but the scalar and pseudo-scalar distributions can
now be apparently distinguished, whereas its still difficult to distinguish
between vector and axial-vector cases.

\begin{turnpage}
\begin{figure*}[hbtp]
\includegraphics[height=0.9\textheight,width=0.33\linewidth,keepaspectratio]{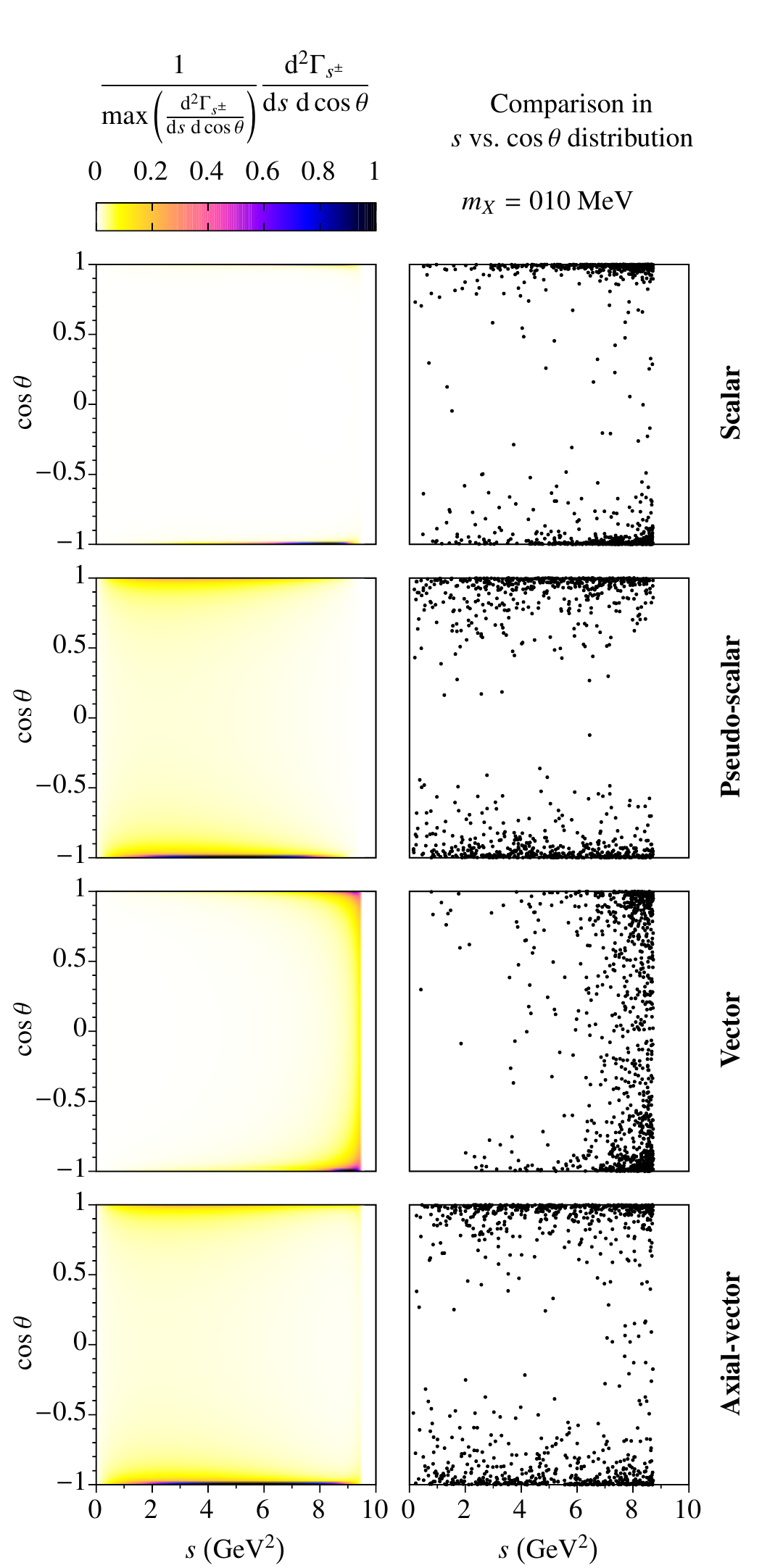}%
\includegraphics[height=0.9\textheight,width=0.33\linewidth,keepaspectratio]{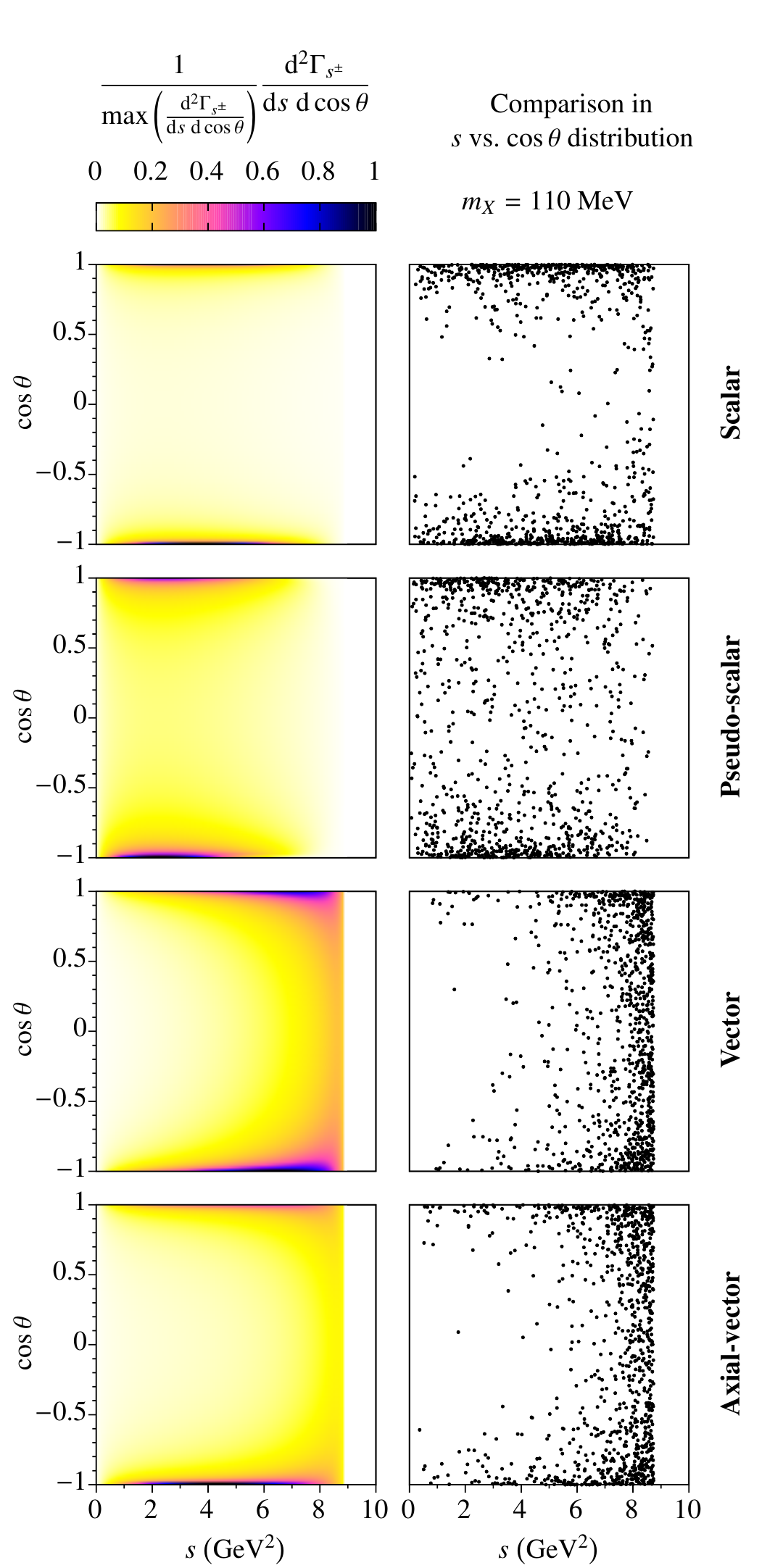}%
\includegraphics[height=0.9\textheight,width=0.33\linewidth,keepaspectratio]{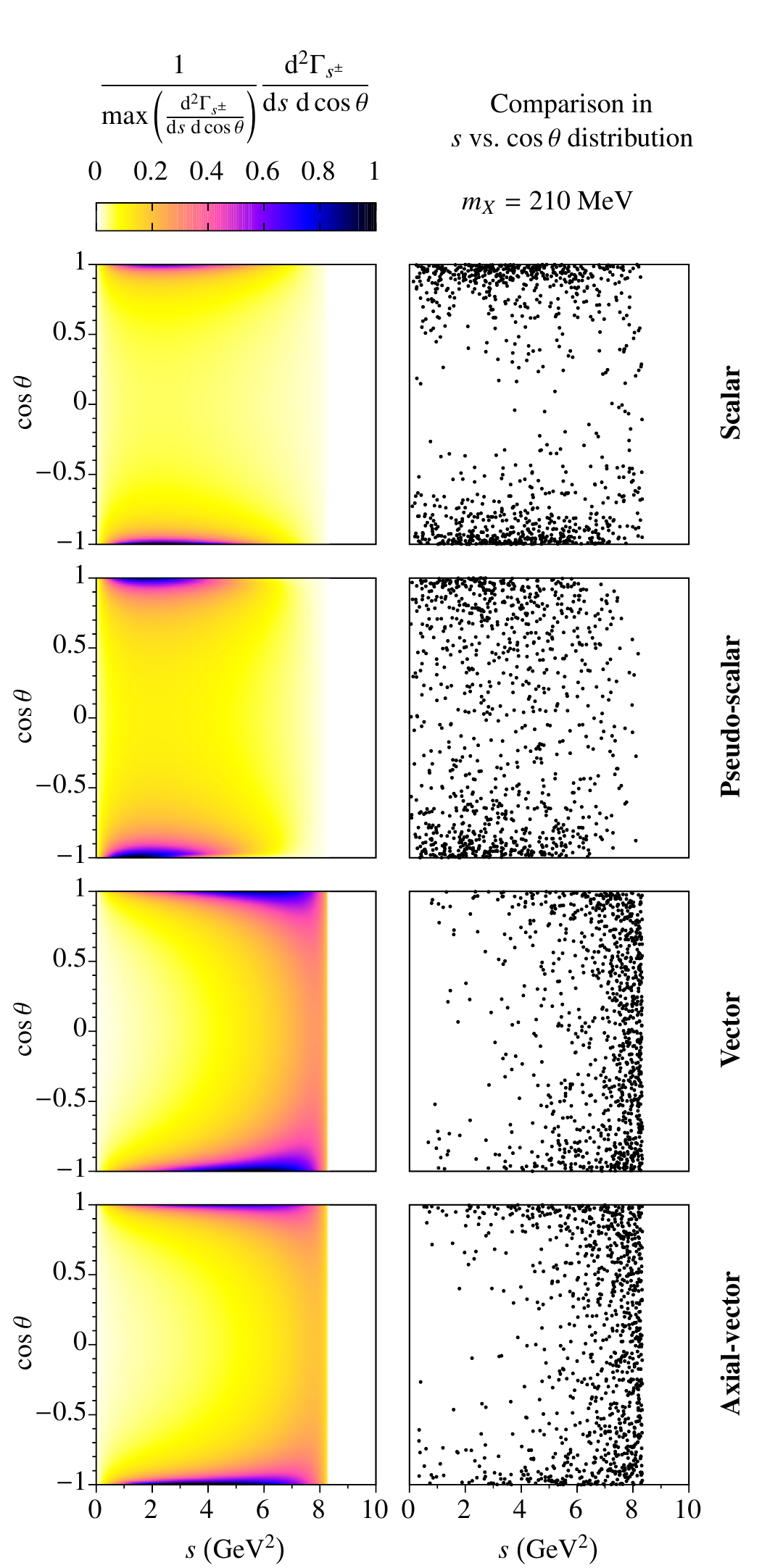}%
\caption{Expected patterns in $s$ vs.\ $\cos\theta$ distributions are shown on
the left panel of each of the subfigures, and on the right panel are the
distributions of $1000$ simulated events after applying a missing energy cut at
$140$~MeV.}%
\label{fig:s-costh-distributions}
\end{figure*}
\end{turnpage}

\paragraph{\bf The \texorpdfstring{$\boldsymbol{s}$}{s} vs.\ \texorpdfstring{$\boldsymbol{\cos\theta}$}{cos(theta)} distribution:} 
Considering the $s$ vs.\ $\cos\theta$ distributions as shown in
Fig.~\ref{fig:s-costh-distributions}, we find that for $m_X=10$~MeV, there are
not enough useful distinguishing features. However, if we consider $m_X=110$~MeV
and $210$~MeV, it is possible to distinguish between all the four spin-parity
cases, albeit the fact that the distribution peaks are still close to the
boundary of the plots.

\begin{turnpage}
\begin{figure*}[hbtp]
\includegraphics[height=0.9\textheight,width=0.33\linewidth,keepaspectratio]{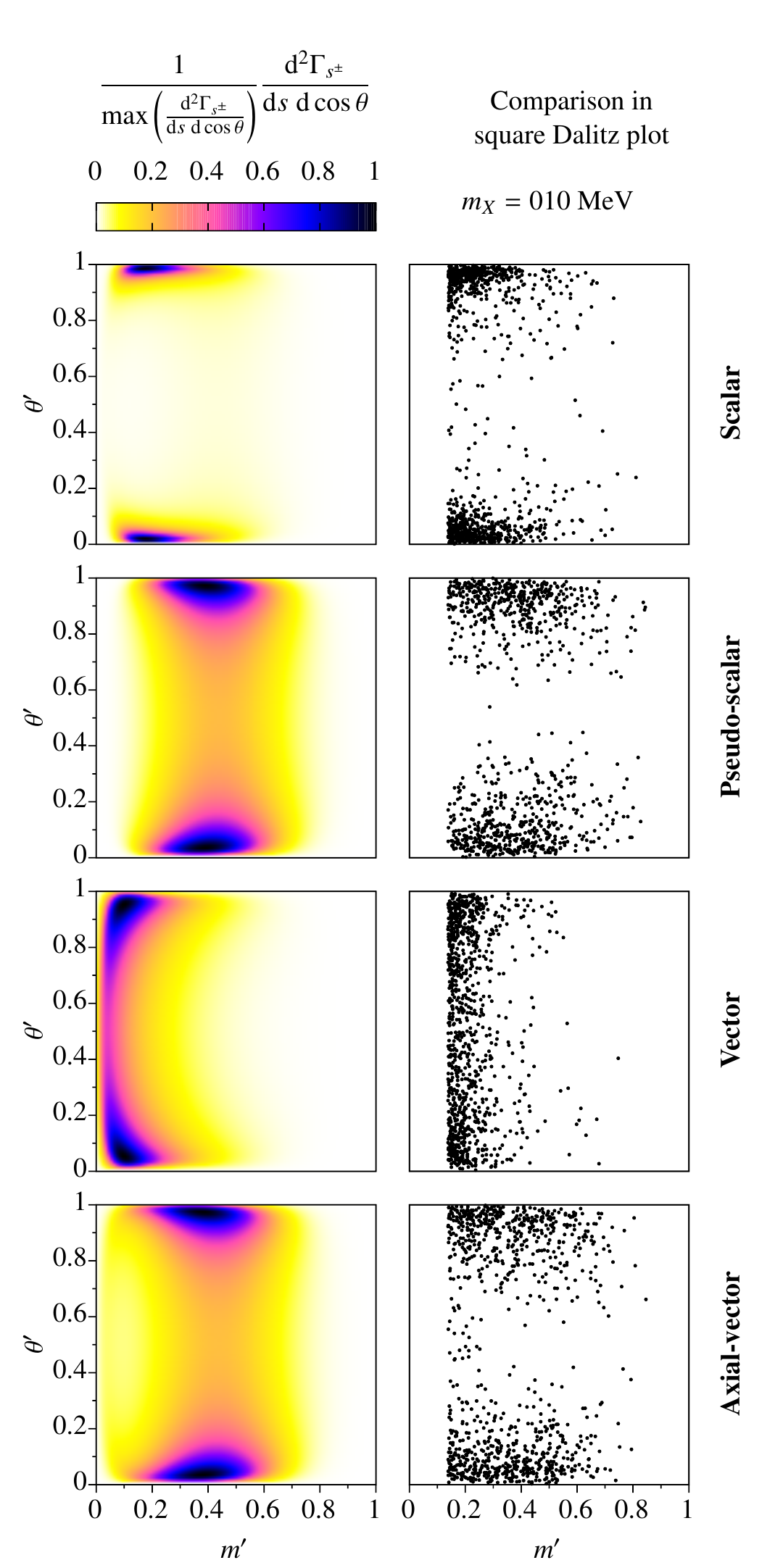}%
\includegraphics[height=0.9\textheight,width=0.33\linewidth,keepaspectratio]{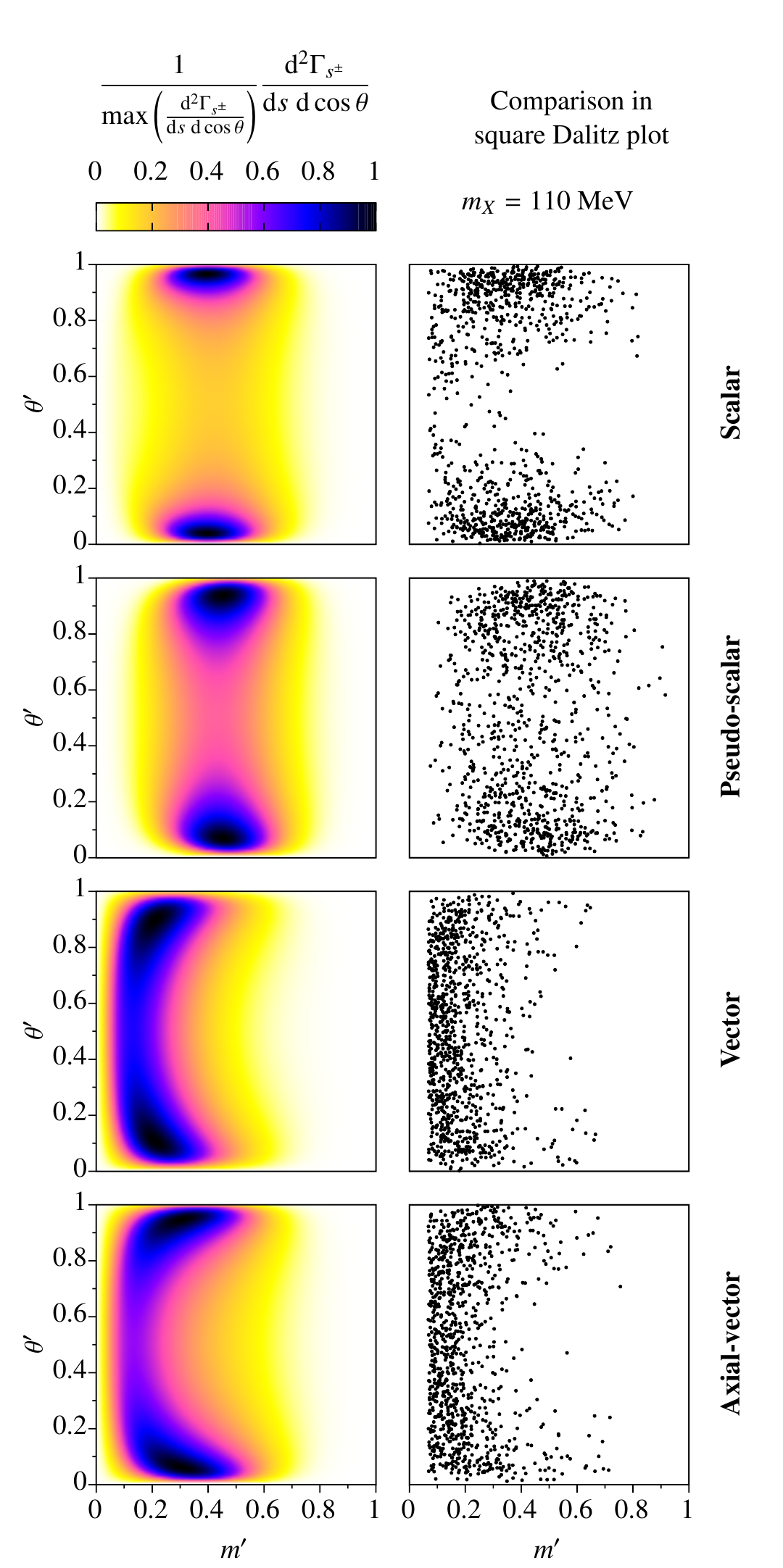}%
\includegraphics[height=0.9\textheight,width=0.33\linewidth,keepaspectratio]{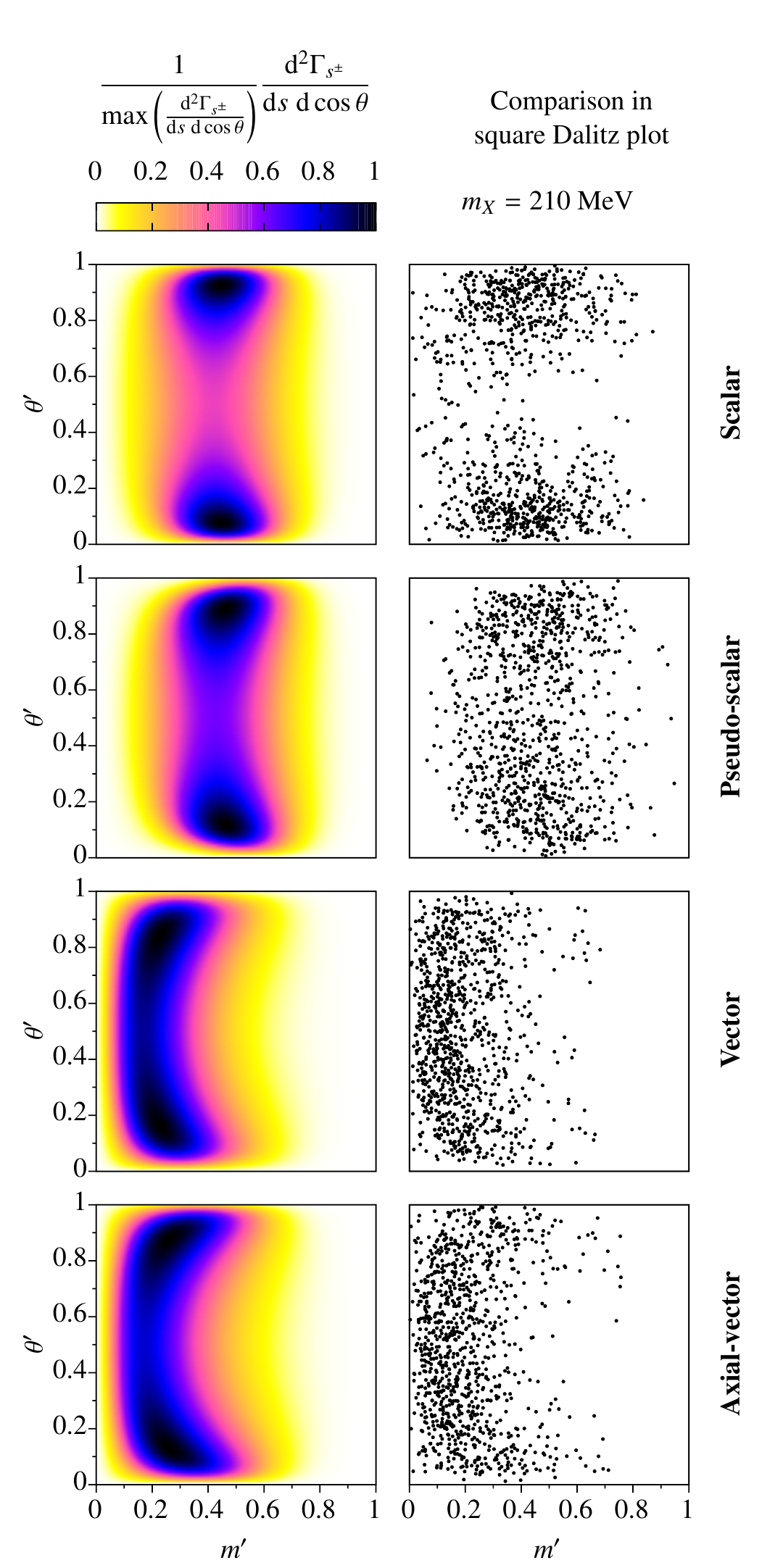}%
\caption{Expected patterns in the square Dalitz plot distributions are shown on
the left panel of each of the subfigures, and on the right panel are the
distributions of $1000$ simulated events after applying a missing energy cut at
$140$~MeV.}%
\label{fig:square-Dalitz-distributions}
\end{figure*}
\end{turnpage}

\paragraph{\bf Square Dalitz plot distribution:}
Considering the square Dalitz plot distributions as shown in
Fig.~\ref{fig:square-Dalitz-distributions} for $m_X=10$~MeV, $110$~MeV and
$210$~MeV, we find that the differences between all the four spin-parity
possibilities which were kind of hidden in other distributions, do show up very
clearly. In fact for the low mass of $m_X=10$~MeV the differences are much
clearly visible, when we compare, for example, the patterns of vector and
axial-vector cases of $m_X=210$~MeV. The square Dalitz plot is indeed a very
powerful discerning distribution due to the fact that the distribution peaks
which are located at the boundaries in other distributions correspond to
internally observed peaks here.

\begin{figure*}[hbtp]
\centering%
\includegraphics[height=0.325\textheight,keepaspectratio]{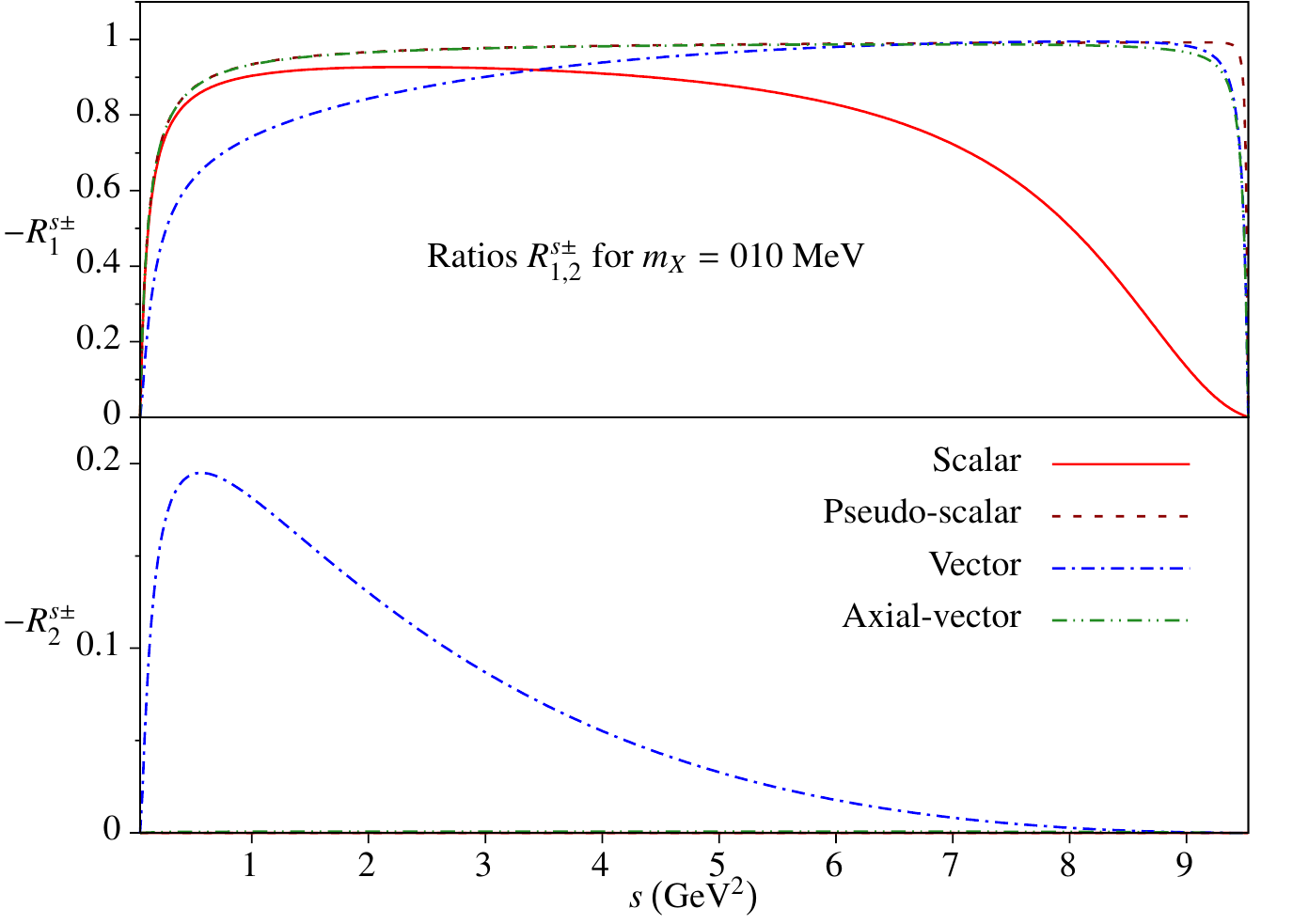}\\[2mm]%
\includegraphics[height=0.325\textheight,keepaspectratio]{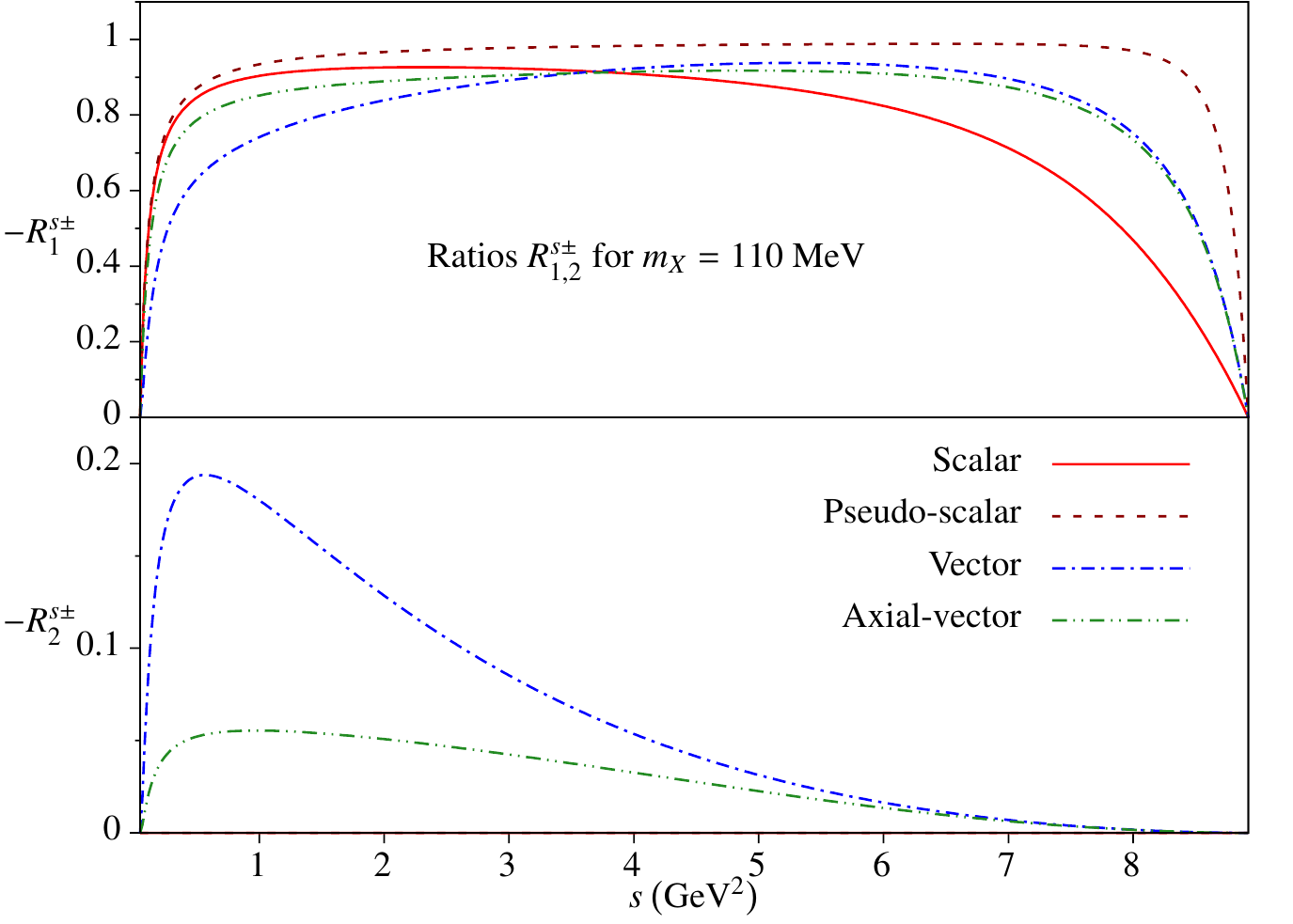}\\[2mm]%
\includegraphics[height=0.325\textheight,keepaspectratio]{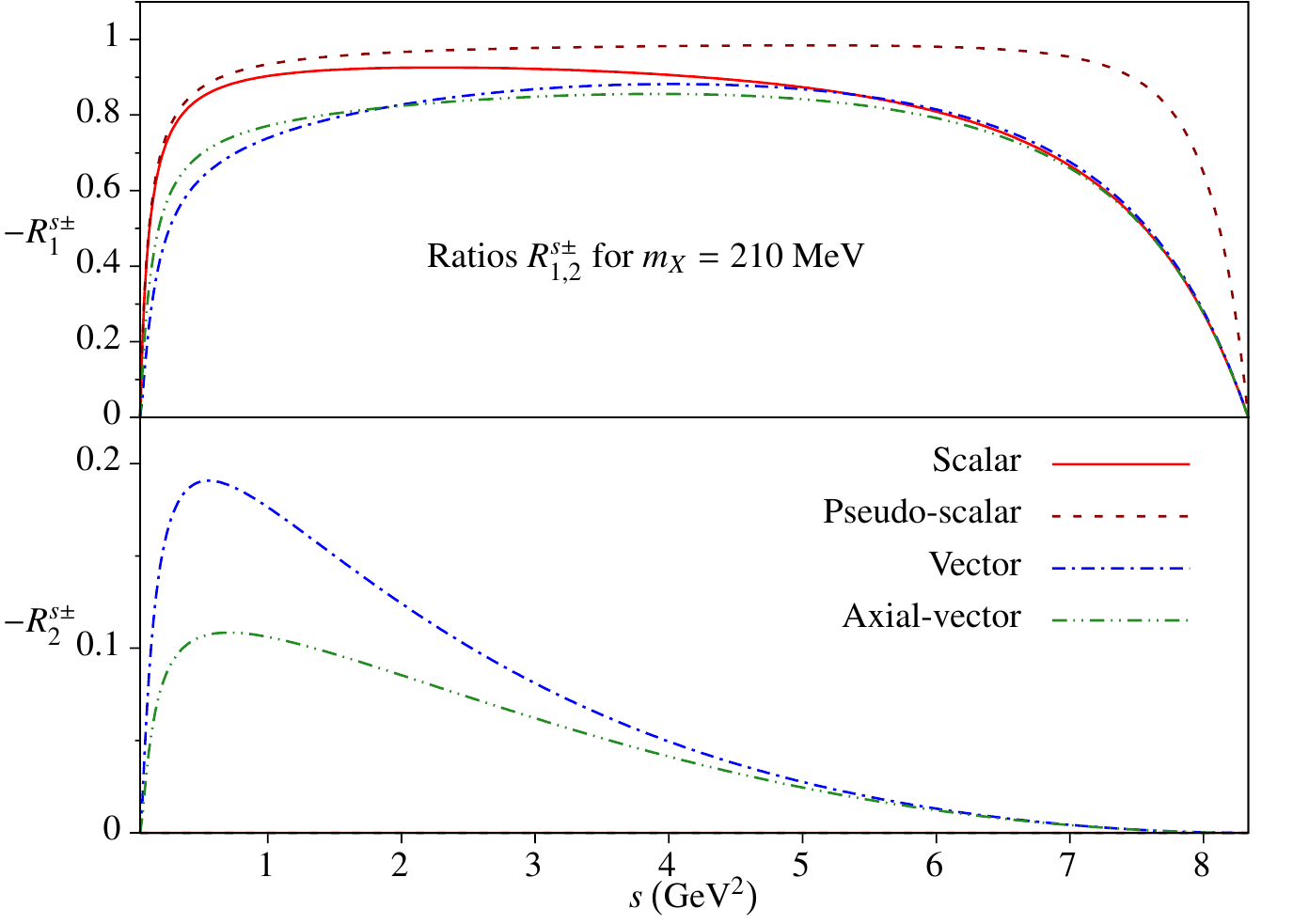}%
\caption{The variation of the ratios $\mathcal{R}_{1,2}^{s\pm}$
for $m_X=10$~MeV with respect to the invariant mass-square $s$.}
\label{fig:Ratios}
\end{figure*}

\paragraph{\bf Ratios:}
Finally looking at the ratios $\mathcal{R}_{1,2}^{s\pm}$, we find from
Figs.~\ref{fig:Ratios} that using $\mathcal{R}_2^{s\pm}=0$ as a criteria one can
easily distinguish the spin-$0$ and spin-$1$ cases. For $m_X=10$~MeV case we
find that for axial-vector scenario the $\mathcal{R}_2^{1-}$ is extremely tiny
but non-zero. In this case one can clearly use the curves for
$\mathcal{R}_1^{1\pm}$ to distinguish the various spin-parity possibilities
without any ambiguity. Therefore, both $\mathcal{R}_1^{s\pm}$ and $R_2^{s\pm}$
should be used to distinguish among the four spin-parity possibilities.

It would indeed be more practical to employ both the ratios and the
distributions, especially the square Dalitz plot distribution to decipher the
spin-parity of the $X$ boson in the decay $J/\psi \to \mu^- \mu^+ X$.

\section{Conclusion}\label{sec:conclusion}

The present anomaly in muon anomalous magnetic moment can be considered as a tantalizing hint of the presence of some new physics. The simplest new physics that can contribute involves purely muon-philic interactions and one needs only to hypothesize that there exists a spin-$0$ or spin-$1$ boson (say, $X$) taking part in these interactions. If the $X$ boson is a parity eigenstate, then only scalar and vector possibilities are allowed from the muon anomalous magnetic moment. Interestingly the allowed paramater space for the coupling constants widens if one considers $X$ to be not a parity eigenstate.

Irrespective of the parity of the $X$ boson, and considering its mass $m_X
\lesssim 2m_\mu$ as suggested by the muon anomalous magnetic moment and other
experimental studies, the electrically neutral $X$ boson can be definitively
probed by the decay $J/\psi \to \mu^- \mu^+ X$. This decay not only provides an
excellent avenue to test this specific muon-philic new physics possibility, but
it also can be utilized to ascertain whether $X$ is a parity eigenstate or not,
and if it is a parity eigenstate we can probe both its spin and parity. Due to
the widening of parameter space for parity non-eigenstate scenario, decay with
such a $X$ would have larger branching fraction than the one which is a parity
eigenstate.

Since the decay $J/\psi \to \mu^-\mu^+ X$ is a three-body decay, there are a
multitude of distributions that can be studied. We find that the square Dalitz
plot distribution as well as the angular distribution of events in the
center-of-momentum frame of the muon pair, do indeed exhibit the distinguishing
features among all the four spin-parity possibilities, viz.\ scalar,
pseudo-scalar, vector and axial-vector cases. We also provide two dimensionless
ratios, obtained from the angular distribution, which are also useful to
ascertain the spin-parity of $X$. The search for decay $J/\psi \to \mu^-\mu^+ X$
or similar decay such as $\phi \to \mu^-\mu^+ X$ would constitute an important
strategy in our search for definitive signature of new physics.

\acknowledgments 

M.M.\ acknowledges the financial support from the DST INSPIRE Faculty research
grant (IFA-14-PH-99), and thanks Indo-French Centre for the Promotion of
Advanced Research for the support (grant no: 6304-2). D.S.\ is thankful to
Donghun Lee for discussions on the numerical study.

\appendix
\begin{widetext}
\section{Expressions for \texorpdfstring{$\boldsymbol{\modulus{A_{s^\pm}}^2}$}{|As+-|**2}}\label{app:AspmSq}

The expressions for $\modulus{A_{s^\pm}}^2$ contributing to
Eq.~\eqref{eq:diff-decay-rate-tu} are as given below.

\begin{subequations}\label{eq:ASq}
\begin{align}
\modulus{A_{0^+}}^2 &= \left(T^2 + U^2\right) \left(4 m_\mu^2 -
m_X^2\right) \left(2 m_\mu^2 + m_J^2\right) + TU \left(T+U\right) \left( T+U +
2\left(4 m_\mu^2 - m_X^2\right) \right) \nonumber\\%
&\quad + TU \left( \left(4 m_\mu^2 - m_X^2\right)^2 - m_X^2 \left(4 m_\mu^2 -
m_X^2\right) + 8 m_J^2 m_\mu^2 \right),\label{eq:ASq-S}\\%
\modulus{A_{0^-}}^2 &= -m_J^2\,m_X^2\,\left(T^2+U^2\right)+\left(T+U
\right)^2\,\left(T\,U-2\,m_X^2\,m_\mu^2\right)-2\,m_X^2\,T\,U\,
\left(T+U-m_X^2\right),\label{eq:ASq-P}\\%
\modulus{A_{1^+}}^2 &= 2\left(T^2 + U^2 \right) \bigg( TU - \left(m_J^2
+ 2 m_\mu^2\right) \left(m_X^2 + 2 m_\mu^2 \right) \bigg) - 4 M^2 TU\left(T+U -
M^{\prime 2}\right),\label{eq:ASq-V}\\%
\modulus{A_{1^-}}^2 &= \frac{2}{m_X^2} \Bigg(
2\,T\,U\,\bigg(m_\mu^2\,\left(T+U\right)^2-m_X^2\, \left(M^{\prime
2}-2\,m_\mu^2\right)\,\left(T+U\right) + m_X^2\,\left(2
\,m_\mu^2\,\left(4\,m_\mu^2-3\,m_X^2-2\,m_J^2\right)+\left(m_X^2+m_J
^2\right)^2\right)\bigg)\nonumber\\*%
&\hspace{1cm} + m_X^2\,\left(T\,U+\left(2\,m_\mu^2+m_J^2
\right)\,\left(4\,m_\mu^2-m_X^2\right)\right)\,\left(T^2+U^2\right)
\Bigg),\label{eq:ASq-A}
\end{align}
\end{subequations}

with $T = t - m_\mu^2$, $U = u - m_\mu^2$, $M^2 = m_J^2 + m_X^2
+ 2m_\mu^2$ and $M^{\prime 2} = m_J^2 + m_X^2 - 2m_\mu^2$.

\section{Expressions for angular coefficients \texorpdfstring{$\boldsymbol{\mathcal{T}_{s\pm}}$}{Ts+-}, \texorpdfstring{$\boldsymbol{\mathcal{U}_{s\pm}}$}{Us+-} and \texorpdfstring{$\boldsymbol{\mathcal{V}_{s\pm}}$}{Vs+-}}\label{app:angular-coefficients}

The expressions for the angular coefficients $\mathcal{T}_{s\pm}$,
$\mathcal{U}_{s\pm}$ and $\mathcal{V}_{s\pm}$ appearing in
Eq.~\eqref{eq:angular-dist} are as follows,
\begin{subequations}\label{eq:angular_coefficients}
\begin{align}
\mathcal{T}_{0+} &=
2\,\left(a-m_\mu^2\right)^2\,\left(2\,a^2+4\,m_\mu^2\,a-2\,m_X^2
\,a+10\,m_\mu^4-6\,m_X^2\,m_\mu^2+8\,m_J^2\,m_\mu^2+m_X^4-m_J^2\,m_X
^2\right)\nonumber\\*%
&\quad -\frac{2}{3}\,\left(2\,a^2+4\,m_\mu^2\,a-2\,m_X^2\,a-6\,m_\mu^4-2\,
m_X^2\,m_\mu^2+m_X^4+m_J^2\,m_X^2\right)\,b^2\\%
\mathcal{T}_{0-} &=
4\,\left(a-m_\mu^2\right)^2\,\left(2\,a^2-4\,m_\mu^2\,a-2\,m_X^2
\,a+2\,m_\mu^4-2\,m_X^2\,m_\mu^2+m_X^4-m_J^2\,m_X^2\right)\nonumber\\*%
&\quad -\frac{4}{3}\,
\left(2\,a^2-4\,m_\mu^2\,a-2\,m_X^2\,a+2\,m_\mu^4+2\,m_X^2\,m_\mu^2+
m_X^4+m_J^2\,m_X^2\right)\,b^2,\\%
\mathcal{T}_{1+} &=
-\frac{24}{15}\,b^4+\frac{8}{3}\,\left(2\,M^2\,a-4\,m_\mu^4-2\,
M^2\,m_\mu^2-2\,m_X^2\,m_\mu^2-2\,m_J^2\,m_\mu^2-M^2\,M^{\prime 2}-m_J^2\,m_X^
2\right)\,b^2 \nonumber\\*%
&\quad +8\,\left(a-m_\mu^2\right)^2\,\left(a^2-2\,m_\mu^2\,a
-2\,M^2\,a-3\,m_\mu^4+2\,M^2\,m_\mu^2-2\,m_X^2\,m_\mu^2-2\,m_J^2\,m_\mu^2+M^2\,M^{\prime 2}-m_J^2\,m_X^2\right),\\%
\mathcal{T}_{1-} &= -\frac{4}{5}\,m_X^2\,b^4
-\frac{4}{3}\,b^2\,\bigg(4\,m_\mu^2\,a^2-8\,m_\mu^4\,a+4\,m_X^2\,m_\mu^2\,a-2\,m_X^2\,M^{\prime 2}\,a+4\,m_\mu^6-4\,m_X^2\,m_\mu^4+2\,m_X^2\,M^{\prime 2}\,m_\mu^2\nonumber\\*%
&\hspace{3cm}
-4\,m_X^4\,m_\mu^2-8\,m_J^2\,m_X^2\,m_\mu^2+m_X^6+3\,m_J^2\,m_X^4+m_J^4\,m_X
^2\bigg) \nonumber\\*%
&\quad +4\,\left(a-m_\mu^2\right)^2\,\bigg(4\,m_\mu^2\,a^2 + m_X^2\,a^2 -
8\,m_\mu^4\,a + 2\,m_X^2\,m_\mu^2\,a - 2\,m_X^2\,M^{\prime 2}\,a + 4\,m_\mu^6 +
13\,m_X^2\,m_\mu^4+2\,m_X^2\, M^{\prime 2}\,m_\mu^2\nonumber\\*%
&\hspace{3cm} - 8\,m_X^4\,m_\mu^2+m_X^6+m_J^2\,m_X^4+m_J^4\,m_X^2\bigg),\\%
\mathcal{U}_{0+} &=
-\frac{4}{3}\,\left(2\,a^2+4\,m_\mu^2\,a-2\,m_X^2\,a-6\,m_\mu^4-2\,m_X^2\,
m_\mu^2+m_X^4+m_J^2\,m_X^2\right)\,b^2,\\%
\mathcal{U}_{0-} &=
-\frac{8}{3}\,\left(2\,a^2-4\,m_\mu^2\,a-2\,m_X^2\,a+2\,m_\mu^4+2\,m_X^2\,
m_\mu^2+m_X^4+m_J^2\,m_X^2\right)\,b^2,\\%
\mathcal{U}_{1+} &=
\frac{16}{3}\,\left(2\,M^2\,a-4\,m_\mu^4-2\,M^2\,m_\mu^2-2\,m_X^2\,m_\mu^2-2\,m_J^2\,m_\mu ^2-M^2\,M^{\prime 2}-m_J^2\,m_X^2\right)\,b^2 - \frac{32}{7}\,b^4,\\%
\mathcal{U}_{1-} &= -\frac{16}{7}\,m_X^2\,b^4 -
\frac{8}{3}\,\bigg(4\,m_\mu^2\,a^2-8\,m_\mu^4\,a+4\,m_X^2
\,m_\mu^2\,a-2\,m_X^2\,M^{\prime 2}\,a+4\,m_\mu^6-4\,
m_X^2\,m_\mu^4+2\,m_X^2\,M^{\prime 2}\,m_\mu^2\nonumber\\*%
&\hspace{3cm} -
4\,m_X^4\,m_\mu^2-8\,m_J^2\,m_X^2\,m_\mu^2+m_X^6+3\,m_J^2\,m_X^4+m_J^4\,m_X^2\bigg)\,b^2,\\%
\mathcal{V}_{0+} &=0,\label{eq:VtermS}\\%
\mathcal{V}_{0-} &=0,\label{eq:VtermP}\\%
\mathcal{V}_{1+} &=-\frac{64}{35}\,b^4,\label{eq:VtermV}\\%
\mathcal{V}_{1-} &=-\frac{32}{35}\,m_X^2\,b^4.\label{eq:VtermA}
\end{align}
\end{subequations}
\end{widetext}

\end{document}